\documentclass[11pt]{article}

\usepackage[table,xcdraw]{xcolor}
\usepackage[utf8]{inputenc}
\usepackage[margin=2cm]{geometry}
\usepackage{graphicx}
\usepackage{setspace}
\usepackage{amsmath}
\usepackage{amssymb}
\usepackage{url}
\usepackage{longtable}
\usepackage{pdflscape}
\usepackage{hyperref}
\usepackage{xcolor}
\usepackage{tikz}
\usepackage{xcolor}
\usepackage{tabularx}
\usepackage{booktabs}
\usepackage{makecell}
\usepackage{titlesec}
\usepackage[defaultlines=2,all]{nowidow}
\usepackage{caption}
\usepackage{hyperref}
\usepackage{nameref}
\usepackage[margin=2cm,font=footnotesize]{caption}
\usepackage{dcolumn}
\usepackage{rotating}
\usepackage{tikz}
\usepackage{svg}
\usepackage{pdflscape}
\usepackage{verbatim}
\usepackage{multirow}
\usepackage{longtable}
\usepackage{lineno}

\titlespacing{\paragraph}{0pt}{0pt}{1ex}

\usepackage[backend=biber,style=authoryear,citestyle=authoryear,isbn=false,url=false,eprint=false,giveninits=true,maxcitenames=1,uniquename=init]{biblatex}

\AtEveryBibitem{
  \clearfield{note}
}

\makeatletter
\newcommand{\setword}[2]{%
  \phantomsection
  #1\def\@currentlabel{\unexpanded{#1}}\label{#2}%
}
\makeatother

\definecolor{bleucite}{RGB}{34,111,212}
\usepackage{hyperref}
\hypersetup{hidelinks}
\AtEveryCite{\color{bleucite}}

\graphicspath{ {./images/} }

\newcommand{\beginsupplement}{%
        \setcounter{table}{0}
        \renewcommand{\thetable}{S\arabic{table}}%
        \setcounter{figure}{0}
        \renewcommand{\thefigure}{S\arabic{figure}}%
     }

\title{Can social capital remedy structural inequality? Economic mobility in a longitudinal population-scale social network}
\author{Yuliia Kazmina$^{1,*}$, Eelke M. Heemskerk$^1$, Emilia van der Kooij$^2$, Eszter Bokányi$^2$, Frank W. Takes$^2$}

\date{{\footnotesize%
    $^1$University of Amsterdam\\
    $^2$Leiden University\\
    $^*$\href{mailto:by.kazmina@uva.nl}{y.kazmina@uva.nl}\\[2ex]
}}

\providecommand{\keywords}[1]{\textbf{\textit{Keywords: }} #1}
\begin{document}

\onehalfspacing

\maketitle
\abstract{

The promise of equal opportunity is a cornerstone of modern societies, yet upward economic mobility remains out of reach for many. Using a decade of population-scale social network data from the Netherlands, covering over a billion family, school, workplace, and neighborhood ties, we examine how structural inequality and social capital jointly shape economic trajectories. Parental background is a strong early predictor of economic outcomes, but its influence fades over time. In contrast, bridging social capital is what positively predicts long-term mobility, particularly for economically disadvantaged groups. Reducing the dimensionality of an individual's network composition, we identify two key dimensions: exposure to affluent contacts and socioeconomic diversity of one’s network. These are sufficient to capture the core aspects of social capital that matter for economic mobility. Overall, our findings demonstrate that while inherited advantage shapes the starting point of economic trajectory, social capital can powerfully reshape it, especially for the poor.

%\textbf{Word Count:} 
%\verbatiminput{abstract_count.txt}

}

\setlength{\parindent}{0em}
\setlength{\parskip}{0.8em}


\newpage

\section{Introduction}
\label{sec:intro}

%\textbf{Word Count:} 
%\verbatiminput{intro_count.txt}

% economic mobility
People generally strive to enhance their well-being, yet paths to economic success vary significantly. While institutions and policies shape the broader environment in which people live and make decisions, life outcomes can diverge dramatically even under the same institutional conditions \parencite{acemoglu2005institutions, breitbach2023institutional}. Some individuals advance economically and socially, whereas others struggle to do so. This disparity in \textit{economic mobility} raises the fundamental question: why do some succeed in improving their economic situation while others do not?

% it's social capital
A long tradition of social capital literature argues that social networks play a crucial role in shaping who gains access to economic opportunities and economic mobility~\parencite{coleman1988social, coleman1990foundations, granovetter1973strength}. Social networks and the resources embedded within them – collectively referred to as \textit{social capital} – provide much-needed access to information and opportunities that might otherwise remain out of reach \parencite{lin2001social, bourdieu1986forms, putnam1993making}. Such benefits most notably emerge when networks connect people from diverse backgrounds, exposing individuals to perspectives and resources beyond their usual social circles \parencite{putnam2000bowling, granovetter1973strength}. This kind of diversity of exposure or ``bridging'' social connectivity have been identified as the most relevant form of social capital that enables economic growth at both individual and community level \parencite{muringani2021social, beugelsdijk2009bonding, callois2007towards}. A recent
large-scale social network study confirmed that individuals from low-income backgrounds who live in neighborhoods characterized by strong cross-class social connectivity are more likely to ascend the socioeconomic ladder than those embedded in more homogeneous networks \parencite{chetty2022social, chetty2022social2}.% with respect to socioeconomic status.
% not needed
%This fragmentation of society in socioeconomic bubbles is further reinforced by homophily, the tendency to form ties with others who share similar socio-demographic characteristics, which strengthens the alignment between social networks and socio-demographic attributes \parencite{Blau1977, McPherson2001, KossinetsWatts2009}. 
%There are systemic biases related to socioeconomic status, gender \parencite{CollischonEberl2021}, race \parencite{PagerShepherd2008}, or ethnicity \parencite{LanceeDronkers2011} that restrict access to high-quality jobs, mentorship, and professional networks, further reinforcing socioeconomic disparities. 
% The social capital literature convincingly shows that social networks contribute meaningfully to socioeconomic outcomes \parencite{GuisoSapienzaZingales2016, KnackKeefer2016, AlganCahuc2014}. 

% we can't have it all / all have it
But are these kinds of favourable social networks available for all? Existing work on \textit{structural inequality} argues that they are not, because access to resources, opportunities, and privileges is systematically organized in ways that perpetuate advantages for some groups --- as defined by a combination of their socio-demographic characteristics --- while creating barriers for others \parencite{pattillo2013black, dimaggio2012network, toth2021inequality}. Social capital, as one of such types of resources, is deeply intertwined with economic and cultural capital \parencite{bourdieu1986forms}, often reinforcing existing societal hierarchies. Because of it, individuals from privileged backgrounds are more likely to access and leverage high-quality networks, perpetuating their socioeconomic advantages. On the other hand, those born into economically disadvantaged households are more likely to attend underfunded schools, limiting their educational attainment as well as their ability to accumulate diverse social capital, further narrowing their future economic prospects \parencite{BourdieuPasseron1977, bourdieu1984distinction, LanceeDronkers2011}. 
% perhaps in the paragraph above making explicit that structural inequality - socionocmomic background = characteristics
%%______________
% existing text

% combined effect unknown! a problem!
This raises a critical question: how do social capital and structural inequality interact in shaping economic mobility? While each has been extensively studied in isolation, their combined effect --- when examined together within a shared institutional context over time --- remains poorly understood. This leads us to a central inquiry: to what extent do social networks complement or pale in comparison to the impact of socioeconomic background?
Addressing this question requires navigating three common challenges in this line of research \parencite{Claridge2021}. The first is the lack of granular data that would comprehensively describe economic outcomes, socio-demographic attributes, and one's social network over time. Even the most advanced studies are conducted at an aggregate, typically, neighborhood or zip-code level, which averages out critical deviations \parencite{chetty2022social, hoogerbrugge2018neighborhood, mohnen2011neighborhood}. Specifically, aggregation hides the nuanced differences in individual experiences and the subtleties of how social networks and socio-demographic attributes interact within these contexts to trigger economic mobility \parencite{TubergenVolker2015, Volker2020}. 

% measuring social capital
Second, there is a challenge of meaningfully translating all information encoded in social networks into metrics of social capital. Social capital is a multifaceted concept that encompasses the strength of close-knit relationships and the breadth of connections across different social groups, as well as the diversity of contacts, among other aspects that may determine the ``value'' of social networks. A key task, then, is to identify which aspects of social capital are most relevant in a given setting and to determine how many distinct dimensions are needed to capture its effects. However, current methods often rely on unidimensional measures of social capital, potentially failing to capture its full impact \parencite{Fine2001, engbers2017}. Such oversimplification also limits our ability to uncover the nuanced ways in which the type of social connectivity that matters for economic mobility may depend on where you are in the socioeconomic distribution. For instance, while bridging ties with more affluent individuals might be crucial for economic mobility among those from disadvantaged backgrounds, bonding ties within one’s immediate community could play a more significant role in maintaining socioeconomic status (SES) for those in middle or upper socioeconomic strata.

The third challenge lies in the persistent difficulty of establishing causal relationships in social capital research \parencite{Mouw2006, BRANCHI20211}. Although certain aspects of social capital, such as connectedness to more affluent groups, are consistently associated with improved economic mobility, the direction of this relationship remains unclear. It is equally plausible that economic mobility enables individuals to form advantageous social ties, rather than social capital being the driver of mobility.

\begin{figure}[!ht] 
    \centering
    \includegraphics[width=\textwidth]{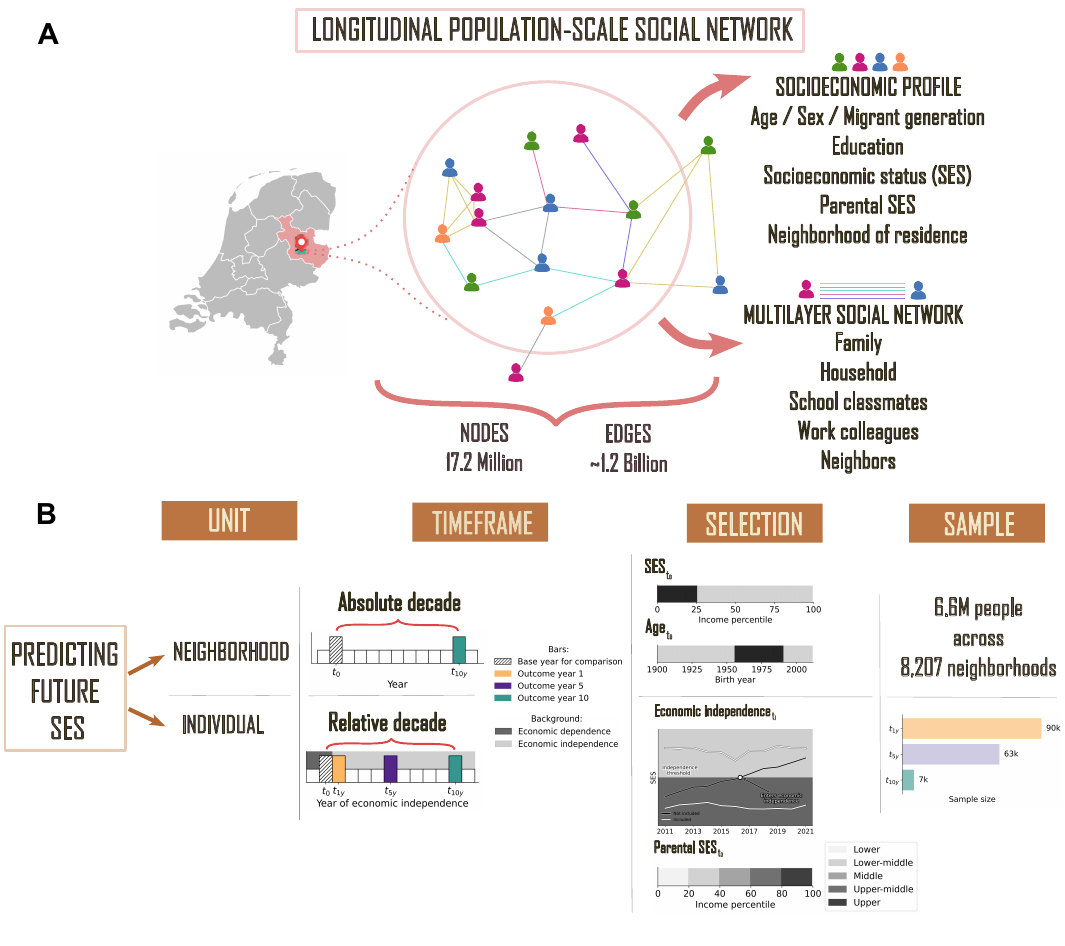}
    \caption{\textbf{Overview of the methodological framework:} (A) Elements of the population-scale social network data, (B)~Unit of analysis, timeframe, and selection pipeline.}
    \label{fig:fig0_results}
\end{figure}

% contribution to overcome challenges
To overcome the aforementioned challenges, we utilize a \textit{population-scale longitudinal network dataset} from the Netherlands spanning a decade, from 2011 to 2021, combining detailed socio-demographic data with multi-layer social network information \parencite{Bokányi2023}. The network data covers 1.3 billion family, household, work, school, and neighbor ties each year for about 17 million residents (Figure~\ref{fig:fig0_results}A). We conceptualize this network as a \textit{social opportunity structure} --- a representation of who is socially reachable within everyday life contexts \parencite{Bokányi2023, Kazmina2024, Menyhert2025, kazmina2024contact}.
% could glue this together with previous paragraph 
The granularity of this data enables flexible aggregation, from individuals to neighborhoods, while providing a full picture of the economic makeup of each person’s social network (see Figure~\ref{fig:fig0_results}B). This means we are not limited to only looking at a certain aspect of the economic composition of their social network, such as, for example, the share of ``rich'' friends or the diversity of socioeconomic backgrounds they are exposed to, which are commonly used in the literature. The population-scale network data allows us to move beyond unidimensional indicators and assess how the broader configuration of social ties across socioeconomic strata shapes opportunities for economic mobility. Finally, leveraging such data allows us to contribute to the so-far unresolved and thorny challenge of understanding causal relationships in social capital research \parencite{Mouw2006, BRANCHI20211}. Disentangling whether social capital drives mobility or vice versa requires longitudinal data that tracks individuals’ socioeconomic trajectories and their evolving social networks over time. Our data enables precisely this: tracking mobility over time while assessing the influence of different forms of social capital, controlling for a rich set of individual-level socio-demographic characteristics. 

In this paper, we adopt a three-step approach. We begin by analyzing the relationship between economic mobility and social capital at the aggregate neighborhood level. 
This allows us to compare and align our findings on the population-scale social network of the Netherlands with recent work on aggregated social media data from the United States~\parencite{chetty2022social, chetty2022social2}, providing a cross-contextual foundation for more granular analysis. Next, we shift our focus to an individual-level analysis, tracking individual life trajectories and economic mobility between 2011 and 2021. This step unpacks the aggregate patterns observed at a neighborhood level and assesses how different types of social capital shape personal economic outcomes over time. Finally, we move beyond traditional measures of social capital. Rather than relying on theory-driven assumptions about which aspects of social networks matter, we reveal the \textit{latent dimensions of social capital} by employing a principal component analysis to the full distribution of socioeconomic status across individuals’ social ties.
This allows us to examine how unique configurations of social connectivity within the population-scale network, rather than broad averages, shape economic outcomes, and which aspects of social capital are consequential for economic mobility depending on individuals’ position in the socioeconomic hierarchy.

Through this multi-level approach on high-quality population-scale data, our study not only tackles the methodological and conceptual challenges encountered in previous work, but also provides a deeper understanding of how structural inequality manifested along socio-demographic lines and social capital interact to shape economic mobility. We find that socio-demographic characteristics independently account for a substantial portion of the variation in economic mobility, underscoring the foundational role of structural inequality in shaping economic mobility. The influence of parental socioeconomic status and other socio-demographic factors on one's economic results is strongest at the onset of one's economic independence, but fades over time as individuals age and accumulate their own experiences and social resources. We also find support for the measurable economic benefits of bridging social capital that not only persist but grow stronger over time. %Through a data-driven approach to social capital, 
Moreover, we identify that there are only two key latent dimensions of social capital that comprehensively describe the socioeconomic composition of individuals’ social networks and matter for economic mobility. These are: (i) the extent to which one is connected to people in the upper half of the income distribution, and (ii) the socioeconomic diversity of one’s social ties. %These findings empirically confirm 
This dual structure aligns closely with longstanding theoretical expectations of what aspects of social capital facilitate economic mobility. % by bridging across social divides. 
These findings offer a robust framework for future research on social capital, validated by high-quality data at the population scale.
%of the relevance of these dimensions for understanding economic mobility.

\section{Results}
\label{sec:results}

%\textbf{Word Count:}
%\verbatiminput{results_count.txt}

The results are presented in three parts: a neighborhood-level investigation of economic mobility, the individual-level determinants of economic mobility, and a data-driven analysis of social capital. Each results subsection is prefaced with a description of the relevant data and methodological steps necessary to interpret the findings, while the full research design is detailed in the \nameref{sec:methods} section.

\subsection{%Insights into s % result sections are always about insights ;) 
Social capital as driver of economic mobility at the neighborhood level}
\label{sec:n_results}
% setup

\begin{figure}[!ht] % move up a page
    \centering
    \includegraphics[width=\textwidth]{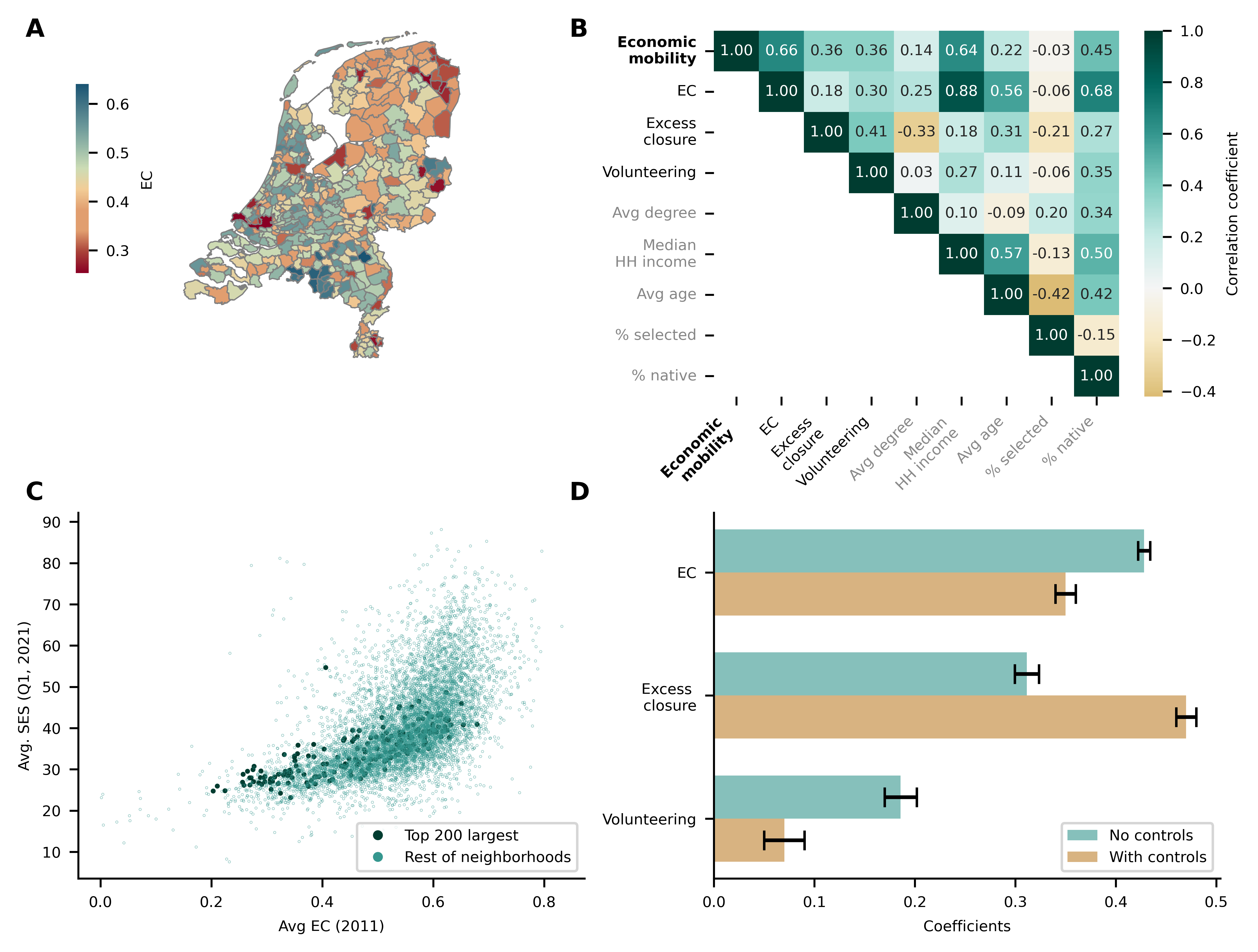}
    \caption{\textbf{Predicting economic mobility at the neighborhood level:} (A) Average EC per municipality in 2011, (B)~Correlation between economic mobility, forms of social capital and neighborhood-level socio-demographic features, (C)~Neighborhood-level average EC or bottom 25\% income distribution in 2011 vs their average income percentile in 2021, (D)~Coefficients and 95\% confidence intervals from a population-weighted multivariate regression predicting neighborhood-level economic mobility as a function of different forms of social capital.}
    \label{fig:fig1_results}
\end{figure}

%Our analysis begins with % storytelling
%predicting economic mobility  of individuals over a period of a decade, between 2011 and 2021. First, 
The first part of our analysis is executed at an aggregated, neighborhood level, where % to establish baseline relationships and situate our findings within the existing literature. In a quasi-replication of 
we follow an approach similar to \cite{chetty2022social}, yet applied to markedly different population-scale data in a different institutional context. After applying the selection criteria detailed in Section~\ref{sec:n_sample} and illustrated in Figure~\ref{fig:fig0_results}B, our final sample consists of 6.6 million residents grouped across 8,207 neighborhoods.
We investigate the chances of those who are at the bottom 25\% of the income distribution to improve their socioeconomic status over the course of 10 years, depending on three types of social capital that characterize their social networks: 
i) social network exposure to those at the upper half of income distribution, i.e., economic connectedness (EC) as defined by \cite{chetty2022social}, capturing bridging social capital, %~\parencite{Peterson2023, choi2023}), 
ii) the level of network closure, capturing bonding social capital  through the measure of excess closure \parencite{Bokányi2023} %, BerardoScholz2010, MussoWeare2015}, 
and 
iii) volunteering rates in the neighborhood, a common expression of bonding social capital~\parencite{Mouw2006,IshamKolodinskyKimberly2003,WilsonMusick1997}.
We furthermore control for a variety of neighborhood-level socio-demographic characteristics. Further methodological details are provided in Section~\ref{sec:neighborhoods_analysis}.  
% ref to methods section for methodological details

% var exploration/visualization
%To contextualize these patterns, we first examine the national distribution of economic connectedness, one of the key predictors in our analysis. 
We find that the distribution of economic connectedness over the whole country is quite heterogeneous (Figure \ref{fig:fig1_results}A). It ranges from ~0.3 to ~0.6, which means that, on average, poor people (who fall in the bottom 25\% of income distribution) who live in these municipalities have somewhere between 30 and 60\% of their social opportunity structure composed of quite economically advanced individuals. Relatively speaking, the north of the country has lower levels of economic connectedness than the south. There is also a difference between urban and rural areas: the largest urban centers, such as Amsterdam, The Hague, Enschede, Maastricht, and Groningen, exhibit the lowest levels of economic connectedness, possibly indicating localized economic isolation.

% var exploration: the 3 social capital operationalizations
To assess the relevance of different forms of social capital for economic mobility, we examine the correlations between each of the three network characteristics and economic mobility (Figure \ref{fig:fig1_results}B). Out of all types of social capital examined: economic connectedness, network closure, and volunteering rates --- economic connectedness is the one that shows the strongest positive correlation with economic mobility. Network closure is positively correlated with economic mobility, but through a weaker relationship compared to EC. Volunteering rates have the weakest correlation with economic mobility among the three measures. 

% PART 1: aggregated level comparison of EC vs mobility, so already longitudinally, variables one by one
Next, to explore in greater detail how economic connectedness relates to economic mobility, contrasting the level of economic connectedness and the socioeconomic status a decade later at a level of neighborhoods %in a scatterplot 
(Figure \ref{fig:fig1_results}C). 
We observe an overall positive relationship between the level of the EC the poorest 25\% have in 2011 and the average socioeconomic status these very same people achieved ten years later.
This relationship is positively linear at lower levels of economic connectedness; however, beyond a certain point, further increases are associated with diminishing or no additional gains in average income percentile. Thus, larger economic connectedness generally supports economic mobility, although we see the marginal benefit diminish beyond a certain threshold.

% and now in a regression
When considered at once in a multivariate regression setting, without the inclusion of any additional controls, economic connectedness at first remains the most significant predictor of economic mobility at the neighborhood level (Figure \ref{fig:fig1_results}D, turquoise bars). This finding corroborates the results of \textcite{chetty2022social}, yet using a different social network data source and in a markedly different institutional and welfare context. 
This underscores the robustness of economic connectedness as a predictor of mobility. In contrast, when we control for socio-demographic characteristics, such as median household income, urbanization, and the share of native residents, the contribution of network closure to economic mobility is higher than that of EC (Figure \ref{fig:fig1_results}D, brown bars). This suggests that the bonding effects of tightly knit networks may play an even more important role in economic mobility than initially apparent, particularly when accounting for baseline neighborhood conditions. In contrast, volunteering rates do not exhibit significant predictive power in this multivariate framework, indicating their limited direct association with economic mobility. 

% some limitations
While the neighborhood-level analysis described above offers valuable insights into aggregate patterns of social capital and economic mobility, it also has important limitations, particularly in addressing questions of causality. By design, this approach aggregates individual experiences into spatial units, which risks conflating compositional effects (who lives in a neighborhood) with contextual effects (what it means to live in that particular neighborhood) \parencite{Kazmina2024}. As such, observed correlations between neighborhood characteristics, such as economic connectedness and economic mobility, may be driven in part by unobserved individual-level factors, such as family background or selective migration.

% PART 2: individual level

\subsection{Individual-level determinants of economic mobility}
\label{sec:ind_results}

\begin{figure}[!ht] % move up a page
    \centering
    \includegraphics[width=\textwidth]{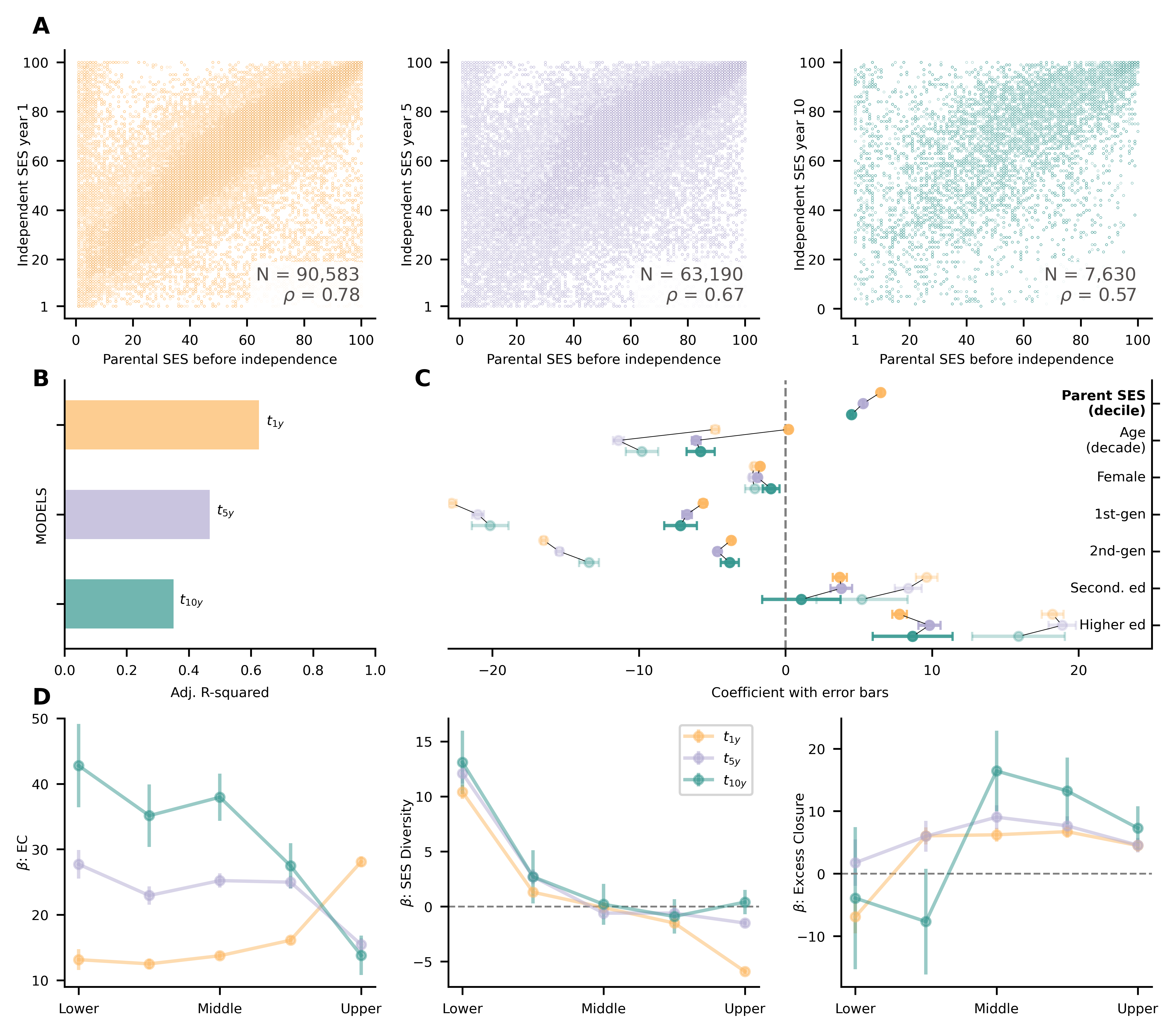}
    \caption{\textbf{Predicting economic mobility at an individual level:} (A) Parental SES before independence vs. individual SES in year 1, 5, and 10, (B) Adjusted $R^2$ when predicting individual SES in years 1,5, and 10 based on individual socio-demographic characteristics and parental SES, (C) Coefficients of multivariate regression predicting individual SES in year 1, 5, and 10 based on individual socio-demographic features (transparent markers) and when controlling for parental SES before entering economic independence (solid markers); full regression results are reported in Table~\ref{tab:table1}, (D) Coefficients and 95\% confidence intervals for forms of social capital (economic connectedness (full regression results are in Table~\ref{tab:table_sc1}), socioeconomic diversity (full regression results are in Table~\ref{tab:table_sc2}), and excess closure (full regression results are in Table~\ref{tab:table_sc3})) predicting economic mobility at years 1, 5, and 10 of economic independence, across five socioeconomic groups.}
    \label{fig:fig2_results}
\end{figure}

To address above mentioned limitations of aggregation, we move to the individual level,  %While still examining the very same relationship -- how different forms of social capital predict people's economic success in the future -- w
where we focus on a specific subgroup of the population: individuals who, for the first time in their lives, enter economically independent lives, no longer relying on support from their parents, the social welfare, or other institutions. The analysis covers the 90,583 individuals who entered economic independence throughout the observation window (see Figure~\ref{fig:fig0_results}B and Section~\ref{sec:i_sample}). We do so because this transitional life stage marks the beginning of individuals’ economic trajectories, and early economic outcomes have been shown to shape future opportunities and compound over time \parencite{blau1967american}.
% the paragraph below could be merged with the previous one 
%Focusing on this group allows us to 

We first examine how individuals’ socio-demographic characteristics, including sex, age, educational attainment, migration background, and, importantly, parental socioeconomic status as an indicator of inherited advantage --- shape their economic mobility at this pivotal moment, using a regression framework that incorporates all these factors as predictors of future SES (for details, see Section~\ref{sec:indiv_analysis}. By isolating these characteristics, we aim to understand how individual traits contribute to economic mobility and %establish a baseline for evaluating the added explanatory power of social capital in subsequent analyses.
%The results of multivariate regression analysis 
provide evidence of the role structural inequalities play in shaping it (Figure \ref{fig:fig2_results}C).

We find that in the first year of their independent lives, economically speaking, women achieve somewhat lower socioeconomic status as compared to men by a few percentage points. However, this disparity appears to nearly close after a decade. Migrants also achieve lower economic results compared to natives. This effect is more pronounced for first-generation migrants: they score more than 20--25 percentage points lower on the SES scale as compared to the natives, for the second-generation migrants, this gap is \~15 percentage points. Given the nature of our sample, which focuses on individuals transitioning from economic dependence (on family or the state) to independence, we exclude first-generation migrants who arrived in the country already economically independent (e.g., labor migrants). Therefore, our conclusions do not apply to this group. Quite intuitively, we also find that entering economic independence with higher levels of education is associated with higher SES in the first year of independence, by \~20 percentage points, however, this advantage weakens over time. On the other hand, individuals who remain dependent on the parental household or the state for an additional decade exhibit lower economic outcomes both five and ten years after eventually making the transition to independence. Altogether, these characteristics of an individual already explain 17\% of the variance in economic mobility (see Table~\ref{tab:table_sc1}). 

Once we incorporate parental socioeconomic status into the model, its predictive power increases to 63\% (Figure \ref{fig:fig2_results}B). A higher socioeconomic status of parents in the year before a person enters economic independence by a decile translates into their own higher SES in the first year of independence by 7 percentage points (Figure \ref{fig:fig2_results}C). This association weakens as we move further into time (Figure \ref{fig:fig2_results}A, C). Five years into economic independence, individual socio-demographic profile and parental SES collectively explain 47\% of the variance in economic mobility (Figure \ref{fig:fig2_results}B). Ten years into independence, with this model, we explain roughly a third of the variance, suggesting that additional factors come into play over time.

Building on this model, which leverages individual socio-demographic characteristics and parental SES as predictors, we look at three distinct forms of social capital relevant at the individual level: (i) economic connectedness (EC), reflecting the prevalence of affluent contacts in an individual’s network; (ii) socioeconomic diversity of exposure (SES Diversity), capturing the spread of SES among an individual’s social ties; and excess closure, measuring the extent of overlap between the social contexts in which individuals are embedded(iii)  (Figure \ref{fig:fig2_results}D).

We apply this extended regression framework that includes socio-demographic characteristics, parental SES and forms of social capital as predictors, separately to five socioeconomic groups, defined by parental SES percentiles prior to economic independence: lower (1-20), lower-middle (21-40), middle (41-60), upper-middle (61-80), and upper (81-100). For each group, we estimate the independent contribution of each form of social capital to future socioeconomic status at three points in time: the first, fifth, and tenth year of economic independence (Figure~\ref{fig:fig0_results}B and Section~\ref{sec:indiv_analysis}). This approach enables us to disentangle how individual background characteristics and social capital jointly shape long-term economic mobility across the socioeconomic spectrum.

From the results presented in Figure~\ref{fig:fig3_results}D (left), we find that economic connectedness has a positive effect across the entire socioeconomic distribution. Regardless of the family background a person is born into, having more economically advantaged individuals in one’s social opportunity structure is associated with better economic outcomes throughout their economically independent years.
Individuals in a lower SES group who are exclusively exposed to people in the top half of the SES distribution achieve higher SES levels than their peers who lack such connections. In the first year of economic independence, this difference is approximately one decile. Over time, the gap widens, reaching about four deciles ten years into economic independence. In contrast, contact diversity appears to have little impact on most individuals, except those in the bottom 20\% of the SES distribution. 

We turn to the other two variables of social capital at the individual level (presented in Figure~\ref{fig:fig3_results}D (center and right)). Socioeconomic diversity of alters for those in the poorest group, in a way, signals the same pattern observed with EC: higher diversity is translated into a higher likelihood of having a more economically advantaged social opportunity structure and is associated with higher economic mobility, by 1 to 1.5 deciles. 
Excess closure contributes positively to economic mobility for nearly all individuals in the first year of independence; however, for those in the lowest SES group, its impact is negative and significant, reducing average outcomes by about 7 percentage points. While this effect on the poorest fades over time, excess closure continues to support improved economic outcomes for the rest of the SES distribution in the longer term.
%\begin{landscape}
While these theoretically informed measures of social capital are well-established in the literature and show clear associations with economic mobility, their specific operationalizations remain somewhat arbitrary. Moreover, it is unclear whether these three dimensions are sufficient to comprehensively capture the structure and value of individuals’ social networks, or whether they reflect redundant aspects of the same underlying phenomena. To address this, we now turn to a data-driven approach that allows us to empirically identify how many and, most importantly, what kinds of latent dimensions are needed to meaningfully describe the social capital embedded in individuals’ networks.

%\end{landscape}
% PART 3: clustering
\subsection{Latent dimensions of social capital shaping economic mobility}
\label{sec:pca_resluts}

\begin{figure}[!b] 
    \centering
    \includegraphics[width=\textwidth]{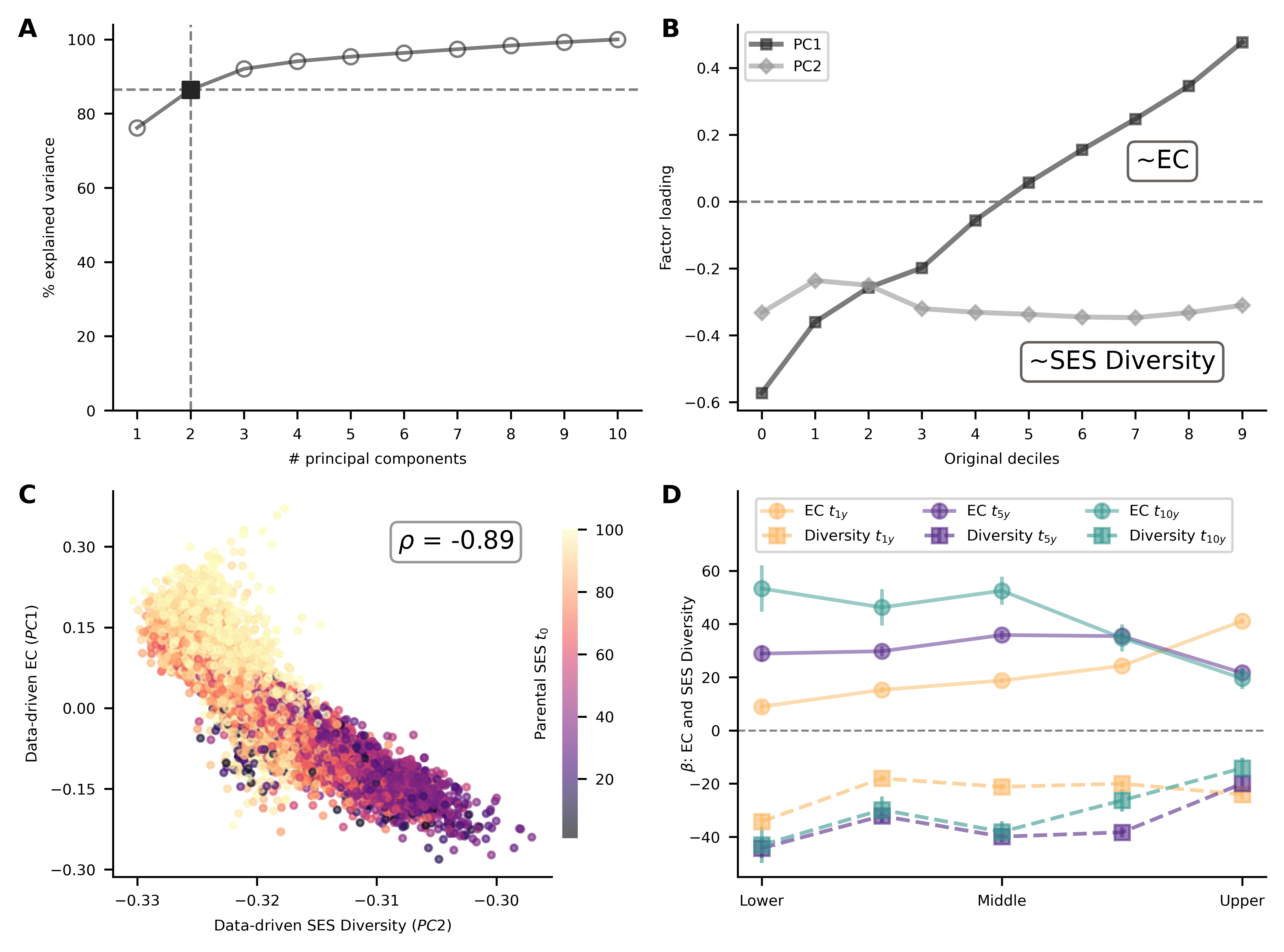}
    \caption{\textbf{Results of PCA on the full socioeconomic composition of individual ties:} (A) Share of explained variance by number of principal components, (B) Factor loadings across SES deciles, (C) Relationship between PC1 and PC2, (D) Coefficients and 95\% confidence intervals for latent dimensions of social capital predicting economic mobility at years 1, 5, and 10 of economic independence, across five socioeconomic groups.}
    \label{fig:fig3_results}
\end{figure}

% TODO adjust to
% composition is what goes IN
% pattern is what comes OUT

% social connectivity pattern = "reserved" term for distribution of SES values of alters of considered ego
In the concluding step of our empirical analysis, we move beyond the three previously employed, theoretically informed definitions of social capital to uncover which aspects of the socioeconomic composition of social networks matter most for economic mobility. To this end, we leverage the full distribution of alters’ SES within each individual’s social opportunity structure in the year prior to their economic independence, and apply principal component analysis (PCA) to identify latent dimensions that provide the most comprehensive description of the composition of one's social networks. 
%To this end, we leverage the full distribution of alters’ SES within each individual’s social opportunity structure, capturing the complete SES composition of their networks. We then apply principal component analysis (PCA) to this socioeconomic composition of social networks to identify latent dimensions of social capital that provide the most comprehensive description of an individual’s social opportunity structure. 
We determine both the number and nature of these latent dimensions, which are then used in place of the theory-driven measures as main predictors in our regression framework (see Section~\ref{sec:pca_analysis} for methodological details). 
%We then determine how many and what kind of latent dimensions in the composition of social ties are relevant for describing one's social networks. These dimensions are then used to assess how different aspects of social capital relate to economic mobility across individuals with varying socio-demographic profiles and starting positions in the SES distribution incorporated in the same regression framework as main predictors of interest, instead of theoretically informed forms of social capital.

%In the concluding step of the empirical analysis, we move beyond the three previously employed, theoretically informed definitions of social capital. The aim is to uncover aspects of the socioeconomic composition of social networks that matter for economic mobility. We therefore leverage the entire distribution of SES of alters within one's social opportunity structure to explain what aspect of it matters for economic success, depending on one's socio-demographic profile and where in the prosperity distribution the egos are. To do so, we perform a principal component analysis (PCA) of the socioeconomic composition of social networks within the sample of economically independent individuals. 
% This data-driven approach allows us to uncover latent dimensions in the composition of social ties -- patterns that may capture complex, multifaceted configurations of social capital beyond the standard typologies. 
%These components are then used to assess how different social connectivity profiles are associated with long-term economic outcomes.

We find that, remarkably, the first two principal components together explain nearly 86\% of the variance, indicating that much of the diversity in the socioeconomic composition of social ties can be captured along just two latent dimensions (Figure \ref{fig:fig3_results}A). Figure \ref{fig:fig3_results}B shows that the first component is characterized by positive loadings on the presence of high-SES alters and negative loadings on low-SES alters, closely resembling the theoretically informed concept of economic connectedness. The second component has similar negative loadings across all deciles, capturing the extent to which an individual's social ties are evenly spread across all SES groups. This dimension thus reflects a form of socioeconomic diversity in one's social opportunity structure.

When we compare individual scores for both dimensions, we find that latent dimensions of social capital are strongly and negatively correlated with each other (Figure \ref{fig:fig3_results}C). This suggests that individuals tend to occupy one of two distinct social network composition profiles: either they are embedded in upward-skewed networks dominated by high-SES contacts, or they are exposed more evenly across the socioeconomic spectrum. In other words, network diversity and economic connectedness appear to be mutually exclusive in practice, reflecting different modes of social integration with distinct implications for economic mobility.

Moreover, we find that individuals with high scores on the first principal component, indicating strong economic connectedness, tend to come from more advantaged family backgrounds (Figure~\ref{fig:fig3_results}C). Specifically, their parents already occupy higher positions in the socioeconomic distribution prior to the individual's transition to economic independence. This association underscores the role of inherited advantage not only in shaping economic outcomes directly but also in structuring access to opportunity-rich social networks.

When we incorporate the principal components into the same regression framework as utilized in Section~\ref{sec:ind_results}, we find that the data-driven dimensions echo the patterns identified through theoretically guided measures (Figure \ref{fig:fig3_results}D). The first principal component --- capturing economic connectedness --- is positively associated with individual economic outcomes, and its predictive power increases over time. In contrast, the second component, capturing network diversity, is negatively associated with economic mobility, with its negative contribution amplifying over time. In this sense, the data-driven approach independently rediscovers the very same dimensions of social capital that were theorized and tested earlier in the analysis, reinforcing the robustness and interpretability of our findings.

\section{Discussion}
\label{sec:discussion}

%\textbf{Word Count:} 
%\verbatiminput{discussion_count.txt}

This work set out to uncover the drivers behind one's economic mobility. Leveraging the well-established literature on structural inequality and social capital, we hypothesize that social networks can provide access to valuable resources and information that foster economic mobility. However, who a person is, defined by their socio-demographic profile and what kind of family they were born into (socioeconomically speaking), matters greatly for the extent to which they can translate this social capital into real opportunities. A previous lack of longitudinal granular data that would describe the socio-demographic profile of a person, characterize their social ties, and track their economic outcomes over time has, until recently, made such an analysis impossible.

Studies conducted at the aggregate level consistently find empirical evidence that cross-class social connectivity plays a key role in fostering economic mobility. For instance, a seminal social capital study from the United States by \cite{chetty2022social} demonstrates that individuals from low-income backgrounds are more likely to achieve economic mobility when embedded in neighborhoods characterized by strong cross-class social ties. Applying a similar framework to the Dutch context, we observe a pattern that closely mirrors previous findings: neighborhoods with higher levels of economic connectedness tend to foster better long-term economic outcomes for individuals from economically disadvantaged backgrounds. Notably, we find that large, more urban, and densely populated municipalities tend to exhibit lower levels of such economic connectedness, indicating that opportunities for cross-class interaction may, perhaps counterintuitively, be more limited in these settings. This stands in contrast to the common perception of cities as spaces where diverse groups naturally interact with each other. In fact, urban environments only foster social fragmentation, with individuals more likely to remain within socioeconomically homogeneous networks \parencite{Kazmina2024}.

Nevertheless, when accounting for the socio-demographic profile of neighborhoods, we find that economic connectedness has an effect comparable to that of social cohesion as measured by clustering. This has two implications. Firstly, bridging ties may not be uniquely transformative, and social cohesion may offer comparable benefits. Bonding social capital and cohesion within communities are equally relevant in supporting economic mobility. Secondly, context matters: in a relatively egalitarian society like the Netherlands, where income inequality is lower than in the US, the marginal benefit of bridging ties may be less pronounced, and thus, more cohesive communities can play a larger relative role than expected.

While these aggregate patterns are informative, they raise further questions. Who actually benefits from these favorable neighborhood structures? What are the distributions behind these neighborhood averages we observe? Drawing on the richness of population-scale social network data from the Netherlands, we thoroughly assessed the relative contribution of socio-demographic profile and social capital to one's economic outcomes on a very granular, individual level over time. 

While both individual profile and accessible social capital make a difference to one's economic mobility, it is, first and foremost, parental socioeconomic status that emerges as the strongest predictor of future economic outcomes, underscoring the enduring impact of structural inequality across generations. Its effect is most prominent in the first year of economic independence and gradually diminishes over time. %, while still accounting for roughly a third of the variation in one's economic outcomes, even a decade after individuals have become economically independent from their parents.
Our findings further reveal that structural inequality is reflected in the varying economic outcomes associated with individuals’ socio-demographic profiles. Characteristics such as educational attainment, migration background, and sex influence economic trajectories not only directly but also by shaping access to and the benefits derived from social capital. These findings highlight that social networks do not operate in a vacuum; rather, their value is conditioned by who the individuals are and the structural positions they occupy.

With regard to specific forms of social capital, we find confirmation that cross-class economic connectedness helps individuals achieve better economic outcomes. In addition, diversity of exposure is also beneficial, but only for those at the bottom of the SES distribution by offering them connectivity to higher-SES individuals that may not exist in more homogeneous networks. Most importantly, while the contribution of social capital to economic mobility is positive and significant in the first year of economic independence, its influence grows steadily over time, becoming even more pronounced a decade into independence.

The data-driven part of the analysis addresses uncertainty about how many and what kinds of dimensions of social capital are necessary to meaningfully capture features of individuals’ social networks that shape economic mobility.
%reinforces the same insights from a different angle. 
Using principal component analysis on the full socioeconomic composition of social networks to uncover latent dimensions of social capital, we find that just two dimensions are sufficient.
These two dominant axes align very closely with long-standing theoretical dimensions in the social capital literature: one reflecting cross-class economic connectedness and the other capturing the socioeconomic diversity of ties. That these dimensions emerge organically from the structure of the data, without imposing predefined categories, underscores their empirical robustness and conceptual relevance.

These findings offer valuable insights into the nuanced and conditional nature of social capital, but they also underscore the importance of methodological precision in capturing these dynamics. Understanding how, for whom, and when social capital matters requires not only theoretically grounded measures but also high-resolution data that can trace individual life courses and social contexts over time. As such, our study contributes to the growing recognition that both the quality of data and the level of analytical granularity are critical for advancing research in this field. 

%add here comments on causality
%Several features of the chosen 
The research design employed in this work 
strengthens the theorized causal directionality 
of the relationship between social capital and economic mobility. 
%the interpretation of these findings in the direction of causal inference. 
First, we follow the same individuals over a substantial period of time, which allows us to track how their social network at the moment of entering economic independence relate to subsequent economic outcomes. This temporal structure of linking social capital measured at baseline to outcomes measured years later helps rule out reverse causality and supports a more plausible order of events.
Second, the relationships we observe are robust to a wide range of individual-level controls, including sex, education, migrant background, and crucially, parental socioeconomic status. Third, we observe consistent patterns across different segments of the SES distribution and over multiple time points in economic trajectories, suggesting these effects are not simply driven by group-specific dynamics or short-term fluctuations.

% limitations
Our study and the data used therein may suffer from limitations.  Unobserved factors, such as professional experience, personality, or unmeasured early-life experiences, may still confound the observed relationships. In addition, our social network measures are derived from administrative data, which, while comprehensive, do not capture the strength or quality of social ties, limiting our ability to distinguish between supportive, neutral, or potentially harmful relationships. 
Yet, recent work also showed that networks from social media platforms such as those used by \cite{chetty2022social} and register-based population-scale social network data such as ours, is strikingly similar in terms of connectivity and structure~\parencite{Menyhert2025}. 
Overall, the strength, consistency, and temporal sequencing of the results lend support to the hypothesis that social capital, and especially economic connectedness, plays a meaningful role in shaping economic mobility trajectories. 
% give it another spin, maybe

% future work
Future research could focus on strengthening the causal understanding of the effects of social capital on economic mobility. To do so, one could leverage natural experiments, policy changes, or instrumental variable strategies that exploit exogenous variation in social network structure or exposure to different forms of capital. Another promising direction lies in unpacking the mechanisms through which social networks influence economic outcomes. Future work could explore the specific pathways that make certain networks more ``useful'' than others, for example, by examining the roles of trust, shared norms, mentorship, reciprocity, or the actual flow of information and resources. 

Ultimately, our findings underscore that fostering economic mobility requires addressing both structural inequalities and the social conditions that shape access to opportunity, recognizing that social capital, while powerful, is neither uniformly accessible nor equally transformative for all.

%To do so, we focus on a cohort of individuals transitioning into economic independence, a critical life stage when long-term trajectories begin to take shape. 

%This approach allows us to isolate the conditions under which social capital is most influential and to assess how personal background, including parental SES, education, migration background, and gender, interacts with different forms of social exposure to shape economic outcomes over time.

\newpage
\section{Methods}
\label{sec:methods}

In this section, we present our methodology and experimental setup. First, we describe the data sources in Section~\ref{sec:data_sources}. Second, in Section~\ref{sec:operationalization}, we outline the variables, including the outcome variables capturing economic mobility and the explanatory variables, both at the individual and neighborhood level. We then elaborate on the analytical choices for creating the individual-level and neighborhood-level samples in Section~\ref{sec:sample}, and conclude by describing the modeling design in Section~\ref{sec:analysis}.

\subsection{Data sources} 
\label{sec:data_sources}

This study is based on a variety of data available through a service of Statistics Netherlands (CBS) called CBS Microdata, which provides individual-level coverage of the entire population of the Netherlands, that is, people registered in the Personal Records Database (BRP) in a given year as of January 1. The data is derived from multiple different administrative registers that are linked to the same unique pseudonymized citizen service number for each individual. In this study, we use i) the socio-demographic profile and ii) the formal social network of people from the time period 2011 to 2022 structured as a series of annual snapshots, as well as iii) neighborhood-level data on volunteering.

\textbf{Socio-demographic profile.} The attributes considered in this study include the following characteristics of individuals: age, highest achieved level of education, sex, and migration background. We retrieve spatial and socioeconomic information: the place of residence up to the level of an administrative neighborhood each year, and the socioeconomic status using annual household income percentile and household prosperity percentile. The latter is a composite measure of income and wealth of a household established by CBS \parencite{cbs_data_documentation}.

\textbf{Social network data.} For the social network, CBS infers relationships between every resident using administrative records on 1st January each year, leading to a collection of formal ties~\cite{van_der_laan_person_2022_correct}. These formal ties cover a variety of social contexts such as family, school, work, household, and neighbors, forming an unweighted multilayer population-scale social network of social opportunities~\parencite{Bokányi2023}. On average, a network of one year contains 17~million people and 1.3~billion formal social ties.

The five social contexts, which constitute the layers of the network, are detailed below:

\begin{itemize}
    \item \textit{Family ties} are derived from administrative data based on parent-child and partner registers and include relationships such as parents, partners, co-parents, children, siblings, grandparents/grandchildren, aunts/uncles, nieces/nephews and cousins. They are also enriched with step-family and in-law relationships.
    
    \item Two people are considered \textit{household} members if they live together at the same address unit. Typically, households consist of single individuals, partnerships, or family members living together. Institutional households such as orphanages, retirement homes, etc., are excluded from the analysis.

    \item \textit{School} edges include various educational levels: primary, secondary/specialized secondary, vocational, and higher education. Two individuals share a classmate link if they attend the same institution during the same academic year and at the same location, pursuing the same educational track of identical duration.

    \item  For each person each year, \textit{work} ties include all colleagues from every workplace the individual was affiliated with for their main source of income during the calendar year preceding 1st January. For cases when a person had more than 100 colleagues throughout this time period, the data provider linked each employee to a subsample of 100 geographically closest colleagues, determined by their place of residence.

    \item \textit{Neighbors} include individuals residing at the ten geographically closest addresses, without imposing any further radius restriction. Additionally, to ensure sufficient representation of the broader neighborhood, particularly in densely built-up areas with many flats, the neighbors layer also contains 20 randomly selected individuals living within a 200-meter radius of an individual's home address. 
\end{itemize}

It is possible for these different layers to overlap; for example, if two individuals are both colleagues and neighbors, their nodes share two distinct connections: a work tie and a neighbor tie. Such overlapping connections are referred to as \textit{multiplex ties}.

\textbf{Civic engagement.} Data on civic engagement is collected by CBS through surveys \parencite{cbs_survey}. It captures a variety of volunteering types, including health care, political parties, and assisting at primary schools or sports societies. The survey assesses the share of people in each neighborhood who are aged between 18 and 65 and do any type of such volunteering work. Data on volunteering rates per neighborhood were only available for the year 2016 or later. For the purposes of this study, we use the 2016 data as a representative midpoint for civic engagement within our observation period.
%The inclusion of people receiving social benefits rather than being employed allows us to measure economic mobility meaningfully, as transitions from social assistance to employment, or the other way around, represent significant, tangible changes in socioeconomic status. 

\subsection{Variable definitions}
\label{sec:operationalization}

In this subsection, we outline how we utilize the previously described data sources to construct variables for further analysis. As data is provided at the individual level, we first explain our measures of economic mobility and its predictors for individuals in Section~\ref{sec:ind_oper}. %Second, we detail the process of capturing their socio-demographic characteristics and the various forms of social capital accessible to them. 
Based on the individual features, we then detail the aggregation steps to the level of administrative neighborhoods in Section~\ref{sec:neighbor_oper}.

\subsubsection{Individual-level variables}
\label{sec:ind_oper}

\textbf{Socioeconomic status (SES) -- dependent variable.} %To evaluate an individual's economic performance over time, we leverage a range of data capturing their socioeconomic status, along with economic information about their parental household. For each individual in the country, we observe their annual household economic outcomes as well as an indicator of whether that person was economically independent in the year of data collection. 
The socioeconomic status of a person is captured by the household prosperity score. \textit{Prosperity score} is a composite measure of standardized household income and assets. Households are ranked based on the sum of the cumulative share in the total income and the cumulative share in the total assets. Based on this sum, the households are ranked from low to high and divided into 100 groups of equal size, where the first group represents the 1\% of households with the lowest prosperity score, and the hundredth group represents the 1\% with the highest economic prosperity.

We use financial independence as a filtering criterion when assessing individuals' economic outcomes. For those who are not yet financially independent or are still in education with income (if any) below the low-income threshold, we assign the prosperity score of their parental household in a given year. 
% This approach allows us to account for the temporary income fluctuations associated with early adulthood transitions, such as pursuing higher education.
%By distinguishing between financially independent and dependent individuals, we ensure that our analysis provides a more accurate representation of economic well-being across different life stages. 

\textbf{Socio-demographic profile -- explanatory variables.} To measure the impact of structural inequalities and barriers associated with certain socio-demographic traits,  we incorporate a range of individual characteristics that are available in the register data. These include age, sex, highest educational attainment, and migration background.
\begin{itemize}
\item \textbf{Age} corresponds to the number of full years a person achieved by the beginning of the year of observation.
% Controlling for age accounts for life-cycle effects on economic and social outcomes. Age influences various factors relevant to our analysis, such as labor market experience, earnings progression, and educational and career transitions.

\item \textbf{Sex} is included as a binary variable (male/female) based on the official register data. It is necessary to acknowledge that such an approach does not account for non-binary and gender-diverse individuals who may experience distinct socioeconomic challenges.
% , allowing us to examine potential differences in economic outcomes and structural inequalities between men and women. 

\item \textbf{Migration Background} classifies individuals into three categories: first-generation migrants (born in the country with at least one foreign-born parent), second-generation migrants (born abroad with foreign-born parents), and natives (born in the country to native-born parents)~\parencite{CBS_ethn}.
% Under this definition, a person is considered native if they were born in the Netherlands and both of their parents were also born in the Netherlands. This category excludes first- (born in the country with at least one foreign-born parent) and second-generation (born abroad with foreign-born parents) migrants \parencite{CBS_ethn}. This measure has important implications for the observed social network of residents: individuals born abroad often have more limited family ties and weaker local support networks compared to native residents.
% This distinction allows us to analyze disparities in economic outcomes and financial independence across different generational backgrounds.

\item \textbf{Education} is treated as an ordinal variable representing the highest level of formal education attained by an individual in the year of observation. The variable takes the following values: primary education (reference category), general secondary and pre-university education, and higher education. 
% Education is included as a key determinant of economic outcomes and social mobility, influencing access to job opportunities, earning potential, and financial independence. 
\end{itemize}

\textbf{Social network -- explanatory variables.} We leverage the multilayer population-scale social network of the Netherlands (see Figure~\ref{fig:fig0_results}A and Section~\ref{sec:data_sources}) to construct meaningful indicators that reflect individuals' social network structure and composition by capturing different forms of social capital from their social networks.

\begin{itemize}
\item \textbf{Degree ($k_i$)} refers to the number of direct neighbors an individual $i$ has within the social network. These neighbors are summed up over the five layers: household, family, school, work, and neighborhood. 
% In the context of our analysis, degree serves as a basic measure of network size, indicating the breadth of an individual's immediate social contacts. 
% A higher degree may suggest greater access to information, resources, and support, which can enhance economic opportunities and financial well-being. However, the quality and nature of these connections also matter, which is why we complement the degree with additional measures.
%\end{itemize} 

%\begin{itemize} 
\item \textbf{Economic Connectedness ($EC_i$)} refers to the extent to which an individual $i$ is socially connected to others from different socioeconomic backgrounds, particularly to those in higher-income groups. It serves as a key indicator of bridging social capital across class lines. We adapt the definition of economic connectedness from \cite{chetty2022social} and operationalize it as a share of individual's contacts that belong to households with above-median prosperity.

\[
EC_i = \frac{|\{j \in \mathcal{N}(i): P_j > P_{\text{median}}\}|}{k_i}
\]

\noindent Here, 

\begin{itemize}
    %\item \( EC_i \) is economic connectedness for individual \( i \). % this is already what you are defining!
    \item \( \mathcal{N}(i) \) is the set of social contacts of individual \( i \) across all layers.
    \item \( P_j \) is the prosperity score (see above) of contact \( j \in N(i) \) .
    \item \( P_{\text{median}} \) is the median prosperity level in the population.
    \item \( k_i = |\mathcal{N}(i)| \) is the degree (number of contacts) of individual \( i \).
\end{itemize}

\item \textbf{Excess Closure} is a slightly adapted version of the well-known node clustering coefficient, tailored to the particularities of a given population-scale social network \parencite{Bokányi2023}. Excess closure only accounts for triangles coming from at least two different layers around egos, discounting the high level of triadic closure coming from the same layer, such as kinship structure or colleague links.
% captures the extent to which an individual's social network is densely interconnected beyond what would be expected given its size and the distribution of degree within each layer.

\item \textbf{Socioeconomic Diversity ($SES$ $Diversity_i$)} is defined as the standard deviation of prosperity of the alters of an individual. A higher value reflects greater exposure to diverse socioeconomic backgrounds within the social opportunity structure of an individual as compared to its mean.
% Greater diversity of exposure may broaden access to varied resources, information, and opportunities.
%$\sigma = \sqrt{\frac{1}{N} \sum_{i=1}^{N} (x_i - \mu)^2}$

\[
\sigma_i = \sqrt{\frac{1}{k_i} \sum_{j \in \mathcal{N}(i)} \left(P_j - \overline{P_i}\right)^2}
\]

\noindent Here, 

\begin{itemize}
    %\item \( \sigma_i \) is the diversity of prosperity (standard deviation) of the social contacts of individual \( i \).
    \item \( P_j \) is the prosperity of alter \( j \).
    \item \( \overline{P_i} = \frac{1}{k_i} \sum_{j \in \mathcal{N}(i)} P_j \) is the average prosperity of the social contacts of ego \( i \).
    \item \( \mathcal{N}(i) \) is the set of social contacts of individual \( i \).
    \item \( k_i = |\mathcal{N}(i)| \) is the degree (number of contacts) of individual \( i \).
\end{itemize}

% add excess closure
%\item \textbf{Average prosperity distance ($d_i^{\text{prosp}}$)} captures the mean difference in prosperity between an individual and their social contacts. Similarly to the diversity of exposure, this measure indicates how diverse the prosperity of one's contacts is. Nevertheless, unlike the standard deviation it compares the prosperity of each person in a social environment not with the mean of the distribution of prosperity, but with the prosperity of the ego. Another crucial difference is the fact that, unlike standard deviation which looks at absolute difference, average distance keeps the sign of the difference. Higher positive values would indicate, that on average an observed ego is connected to people below their own prosperity group. Negative values capture cases when ego's social environment is more prosperous than the ego themselves. 
%\[
%d_i^{\text{prosp}} = \frac{1}{k_i} \sum_{j \in \mathcal{N}(i)} (P_i - P_j)
%\]
%\noindent where:

%\begin{itemize}
%    \item \( d_i^{\text{prosp}} \) is the average prosperity distance for individual \( i \).
%    \item \( P_i \) is the prosperity of the focal individual \( i \) (ego).
%    \item \( P_j \) is the prosperity of each alter \( j \).
%    \item \( \mathcal{N}(i) \) is the set of social contacts of individual \( i \).
%     \item \( k_i = |\mathcal{N}(i)| \) is the degree (number of contacts) of individual \( i \).
%\end{itemize}

\end{itemize} 

\subsubsection{Neighborhood-level aggregation of individual data}
\label{sec:neighbor_oper}

% Having operationalized the data at the individual level, we now aggregate these individual observations into larger analytical units. This aggregation enables comparisons with existing literature in the field.

The chosen level of aggregation is the smallest available level of administrative neighborhoods, known as a ``buurt'' in Dutch. A typical neighborhood usually consists of roughly 500 to 3000 residents. However, there is substantial variation depending on the density and urban structure. % Urban neighborhoods with higher density would have somewhere between 1000 and 3000 residents. Neighborhoods in suburban areas are homes to 500-2000 people. And rural areas are much less densely populated, with less than 500 residents per neighborhood. Such ``buurts'' typically serve as the finest-grained administrative units available in Dutch population data.

\textbf{SES -- dependent variable.} As described in Section~\ref{sec:n_sample}, economic mobility at the neighborhood level is measured exclusively for individuals who have socioeconomic data that can be meaningfully compared between 2011 and 2021. Specifically, we focus on individuals within each neighborhood who belong to the bottom quartile (Q1) of the household income distribution in 2011. Their household income percentile position in 2021 serves as a proxy for their economic mobility.
The neighborhood-level measure of economic mobility is then computed as the average income percentile in 2021 for residents who belong to the lowest quartile (Q1) of the household income distribution in 2011. This definition aligns closely with the approach introduced by \cite{chetty2022social}, who similarly measure economic mobility as the average income percentile at the second observation point for individuals who were at the 25\textsuperscript{th} percentile of the income distribution at the initial observation.

\textbf{Socio-demographic profile -- explanatory variables.} The list of features describing a socio-demographic profile of each neighborhood is as follows:

\begin{itemize}

\item \textbf{Average age} of residents of a neighborhood captures generational characteristics and is calculated based on all neighborhood residents.

\item \textbf{Share of natives} represents the proportion of individuals in the neighborhood who are classified as native Dutch according to the definition used by CBS and is calculated based on all neighborhood residents. 

\item \textbf{Median household income} captures the middle point of the income distribution among households in the neighborhood and is calculated based on all neighborhood residents.% This measure provides a robust indicator of the neighborhood's overall economic standing in the country's population, reflecting residents' financial resources and access to economic opportunities.
\end{itemize}

% Given the sample construction strategy described in Section~\ref{sec:n_sample}, certain subgroups of the population are excluded from the analysis. Overall, this filtering ensures that the resulting sample comprises individuals with representative socioeconomic data at both the beginning and the end of the observation period. 
To correct for the data artifacts  associated with the sample construction strategy (see Section~\ref{sec:n_sample}), we also control for:

\begin{itemize}
    \item \textbf{Share of selected people}, capturing the proportion of neighborhood residents included in the analytical sample relative to the total population of a neighborhood, which controls for differences arising from sample composition.
\end{itemize}

\textbf{Social network -- explanatory variables.} We aggregate individual-level network measures to reflect the broader social network metrics within neighborhoods: 

\begin{itemize}
\item \textbf{Degree} is calculated as the average network degree of all individuals residing in the neighborhood.
% , capturing the overall level of social connectivity among residents.
\item \textbf{Excess closure} is calculated as the neighborhood-level average of individual excess closure values, calculated based on all neighborhood residents.
\item \textbf{Economic Connectedness} on the neighborhood level was defined in the same manner as \cite{chetty2022social}. We calculate the average share of contacts above median SES among those residents who themselves are below median SES. %This measure reflects the extent to which lower-SES residents are socially connected to individuals from higher socioeconomic backgrounds. 
\item \textbf{Civic engagement} is presented as a share of neighborhood residents aged between 18 and 65 who, according to the survey, do some type of volunteering work. 
\end{itemize}

\subsection{Sample construction}
\label{sec:sample}

%Despite the abundance of available data, we carefully design our sample to ensure that it is consistent and well-suited to the analytical goals of this study. 
Below we describe the strategy for sampling the population for the individual-level analysis (Section~\ref{sec:i_sample}). We then describe the sampling strategy for the aggregated analysis at the neighborhood level in Section~\ref{sec:n_sample}.

\subsubsection{Individual-level analysis}
\label{sec:i_sample}

%For the individual-level analysis, the sample selection differs. 
From the entire population, we focus specifically on individuals who achieve economic independence for the first time at any point within the observation period (2011–2022). A person is considered economically independent if their net income from labor or business is at least 70\% of the legal minimum wage in a given year --- the equivalent of social assistance for a single person \parencite{CBS2018}. 
%While the sample is not explicitly restricted by age, the transition to economic independence typically occurs between the ages of 16 and 28. Figure XX illustrates the distribution of birth years of people who attain economic independence throughout the observation period. Figure XX illustrates the economic role (?) of these individuals the year before they become economically independent. --> SI
%The sample includes individuals across the entire socioeconomic spectrum, regardless of their parental household's prosperity level before they transition to economic independence. Individuals are ranked based on the prosperity level of their parental household prior to attaining economic independence, after which they are categorized into five groups: lower (1-20), lower-middle (21-40), middle (41-60), upper-middle (61-80), and upper prosperity levels (81-100). %The number of people in each prosperity group is presented in Figure~\ref{fig:fig_SI}.
The sample consists of 93,717 individuals representing the full socioeconomic spectrum, defined by the prosperity level of their parental household prior to their transition to economic independence. Due to missing data, primarily in educational information, we retain 90,583 complete observations for analysis. Based on this measure, individuals are ranked and grouped into five categories: lower (1–20), lower-middle (21–40), middle (41–60), upper-middle (61–80), and upper (81–100).

\begin{figure}[!ht] % move up a page
    \centering
    \includegraphics[width=0.5\textwidth]{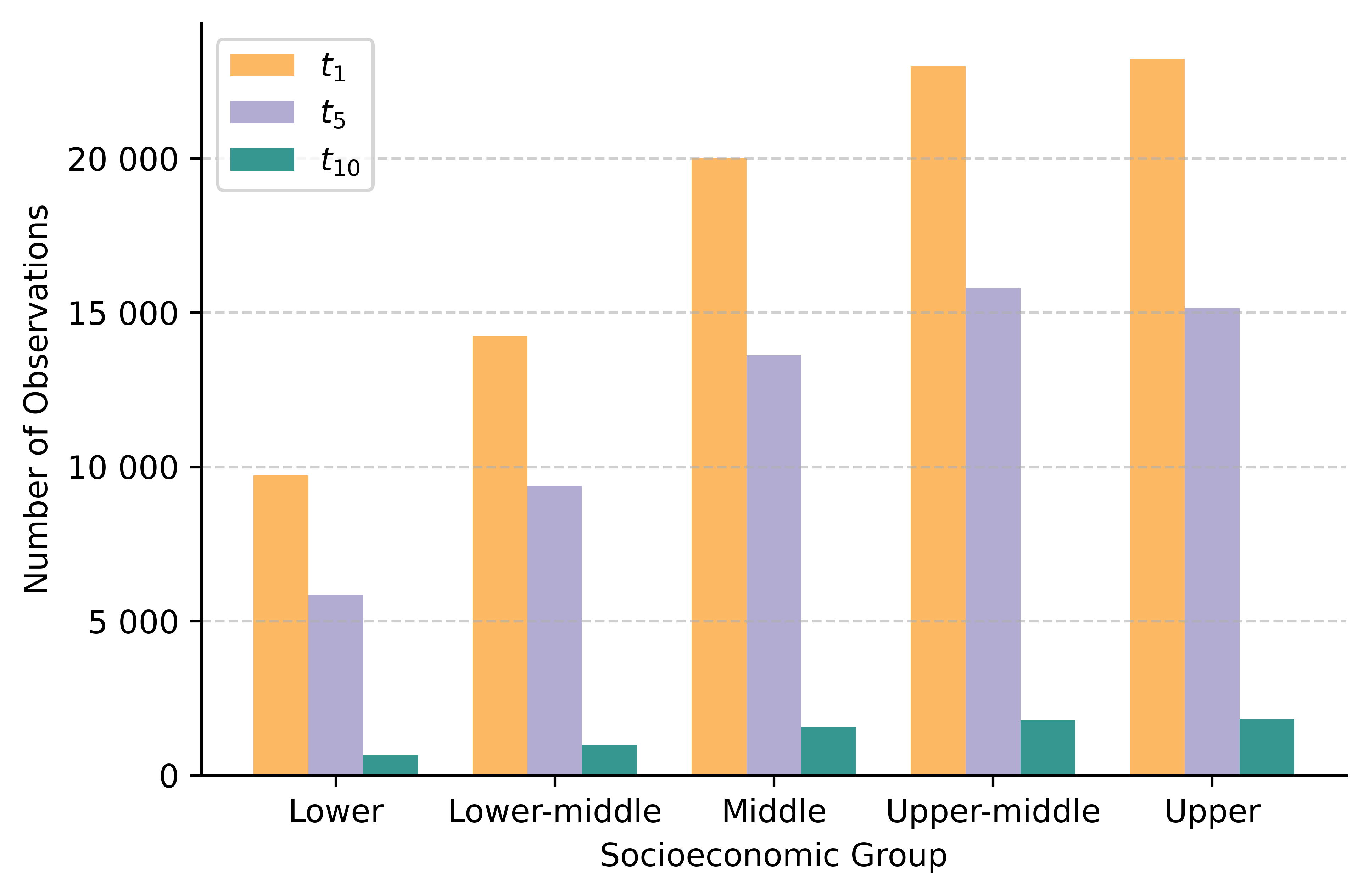}
    \caption{\textbf{Longitudinal composition of the sample:} Number of observations in the sample at years 1, 5, and 10 of economic independence, across five socioeconomic groups.}
    \label{fig:fig_SI}
\end{figure}

%This leaves us with 93,717 individuals, who have differing lengths of observable independent economic trajectories: for individuals who became economically independent in 2021, we capture only one year of their own economic outcomes, whereas individuals who attained independence in 2011 can be tracked across the full 12-year period. %Figure XX presents the distribution of individuals in the sample by the year they achieved economic independence. 
%Figure~\ref{fig:fig_SI} illustrates the distribution of individuals based on the duration for which their economic independence is observed.

The duration over which individuals’ independent economic trajectories are observed varies. Those who became economically independent in 2021 contribute only one year of outcome data, while those who transitioned in 2012 can be followed for up to a decade. Figure~\ref{fig:fig_SI} presents the distribution of individuals in the sample across both the length of observed economic independence (discretized at years 1, 5, and 10) and their parental socioeconomic group.

\subsubsection{Neighborhood-level analysis}
\label{sec:n_sample}

We aim to have a sample of people who have a valid source of income in both 2011 and 2021. Therefore, in the neighborhood-level sample, we include:
\begin{enumerate}
\item people with a known household income in both 2011 and 2021;
\item people who were born between 1955 and 1990. As such, by the beginning of the observation period in 2011, we have people between ages 21 and 56, since 21 is the age at which employees
get to receive full minimum wage, and those who are older 56 in 2011 will most likely retire by the end of the observation period in 2021 (as the majority of people in the Netherlands retire after the age of 66);
\item people with income, as well as people receiving social benefits. Selection is based on the CBS Microdata variable ``socioeconomic situation'', through which we identify and include employees, director-majority shareholders, self-employed entrepreneurs, and recipients of unemployment, social assistance, illness, disability, or other social security benefits. We exclude people receiving pension benefits in either 2011 or 2021, to avoid measuring the effect of retirement on income over time. We exclude students with or without income, because their change in income is likely affected by moving from no job/part-time job to a full-time job. We also exclude people classified as a director-majority shareholder who have missing or incomplete income data. Lastly, we exclude people in
households with no perceived income, as well as members of institutional households (orphanages, prisons, retirement homes).
\end{enumerate}
After applying these filtering steps, our final sample comprises 6.8 million individuals, out of an original population of approximately 16.69~million in 2011. %This sample represents those for whom we have reliable socioeconomic information in both time periods, enabling us to capture genuine economic mobility. 

When linking individuals to their neighborhood of residence, there are 321 individuals for whom this is unknown, so they are further excluded from the sample. We group the remaining observations by the neighborhood of residence. Following the disclosure risk policies of the data provider, we make sure that each neighborhood has at least 10 people in the bottom 25\% of prosperity distribution in 2011 (as this is the primary group of analysis as described in Section~\ref{sec:neighborhoods_analysis}). This step excludes either very small neighborhoods or those who have primarily wealthier residents. After the neighborhood selection, 8,207 of the full sample of 11,547 neighborhoods remain. With this,
another 200,386 people are excluded from the sample, resulting in a final analytical sample of approximately 6.6~million people residing in 8.2k neighborhoods.

\subsection{Statistical analysis} 
\label{sec:analysis}

Below, we start with a comprehensive analysis of neighborhood-level patterns of economic mobility in Section~\ref{sec:neighborhoods_analysis}, followed by an individual-level analysis. This includes a theory-driven examination of key social capital measures in Section~\ref{sec:indiv_analysis} as well as a data-driven approach using principal component analysis in Section~\ref{sec:pca_analysis}.% to uncover latent dimensions of social network composition 

\subsubsection{Neighborhood-level economic mobility}
\label{sec:neighborhoods_analysis}

For the neighborhood-level analysis in Section~\ref{sec:n_results}, we employ ordinary least squares (OLS) models weighted by the number of residents to estimate economic mobility, defined as the average income percentile in 2021 of individuals who were in the bottom half of the income distribution in 2011. All variables are standardized to have a mean of zero and a standard deviation of one to facilitate comparison of effect sizes both within our analysis and with findings from other studies.

The first regression model, that we refer to as ``No controls'' in Figure~\ref{fig:fig1_results}D, incorporates three forms of social capital, namely, EC of those with below-median SES, Excess Closure and Volunteering in neighborhoods as predictors. 
We further include neighborhood-level control variables such as average degree, median household income, average age, share of native-born residents, and the share of individuals included in the analytical sample. We refer to this model as ``With controls'' in Figure~\ref{fig:fig1_results}D.

%Finally, we evaluate the temporal stability of neighborhoods as analytical units, examining whether their composition remains consistent over time. To do so, we calculate the share of individuals who resided in a given neighborhood in 2011 and were still living there in 2021. We display and examine the distribution of this temporal stability measure.

% \end{enumerate}

\subsubsection{Individual-level economic mobility}
\label{sec:indiv_analysis}

The framework of the individual-level analysis in Section~\ref{sec:ind_results} is structurally similar to that of the neighborhood-level analysis. However, greater emphasis is placed on socio-demographic characteristics. In this context, we treat socio-demographic features as key predictors in their own right and interpret their individual contributions to economic mobility.
%these variables are not merely included as controls but are treated as key predictors of interest. 
%This part of the analysis, in the same manner, aims to predict individuals' economic success. This prediction is based on both who individuals are, i.e., their socio-demographic characteristics, and the forms of social capital accessible to them. Beyond making this prediction, our goal is to uncover the relative contributions of these two groups of predictors. As detailed in \ref{sec:i_sample}, we conduct these predictions for individuals at the onset of their economic independence. 
Using information on who they are and the forms of social capital accessible to them prior to this transition, we predict individuals' economic prosperity in the first, fifth, and tenth year following the start of their economically independent lives. The individual-level analysis is conducted in two steps:
\begin{enumerate}
    \item \textit{Socio-demographic predictors of economic mobility.}

    We first show how much parental socioeconomic status matters for one's own economic outcomes. We compare parental socioeconomic status in the year prior to an individual's economic independence with that individual's own prosperity in years 1, 5, and 10 following the transition.
    
    Next, we examine the contributions of individual socio-demographic characteristics, including the age at which a person enters economic independence, their sex, migrant background, and highest level of education attained. We do so by using all these individual socio-demographic features as predictors of individuals' own prosperity in the years 1/5/10 of their economic independence in an OLS multivariate regression. 
    
    Finally, we combine the information on parental SES and individual socio-demographic characteristics as predictors of individuals' own economic outcomes at multiple stages of their economic independence. We also compare the share of explained variance (Adjusted R-squared) such a model has in the first year of economic independence as well as in the subsequent years.

\item \textit{Social capital as a predictor of economic mobility.}

% Using the model that predicts individual economic success based on socio-demographic characteristics and parental SES as a baseline, we then introduce the social network dimension to further explore its explanatory power. Importantly, we conduct this analysis separately for distinct socioeconomic groups. We divide the sample into five categories: lower, lower-middle, middle, upper-middle, and upper prosperity groups-based on parental SES measured one year prior to an individual's transition to economic independence. %For each group, we first add controls for individual social network structure, namely, one's degree and excess closure in a social opportunity structure. 
We incrementally introduce different forms of social capital as additional predictors, assessing their contributions individually. 
We first add EC the year before economic independence as a social capital predictor of SES. 
Alternatively, we incorporate the Socioeconomic Diversity. 
Finally, we incorporate the Excess Closure across various social contexts. 

All social network-based variables are captured a year before a person becomes economically independent. The individual contributions of EC, SES Diversity, and Excess Closure to a person’s independent SES are assessed in the first year of economic independence, as well as in years 5 and 10. 

\end{enumerate}

% Together, these neighborhood- and individual-level analyses provide a comprehensive understanding of how socio-demographic background and different forms of social capital interact to shape economic mobility, both across communities and throughout individual life trajectories.

\subsubsection{Latent dimensions of social capital}
\label{sec:pca_analysis}

To uncover latent dimensions of social capital in Section~\ref{sec:pca_resluts}, we apply a data-driven, unsupervised approach that captures variation in the SES profiles of individuals' social ties without relying on predefined measures such as the share of affluent contacts or the standard deviation of SES.
For consistency, we rely on the same sample of individuals used in Section~\ref{sec:indiv_analysis}. %, who are entering their economically independent lives. 
Each individual is represented by a vector indicating the proportion of their social contacts in each SES decile. These SES distribution vectors were row-standardized to sum to 1, ensuring comparability across individuals.
% refer to SI on more details on these distirbutions, if necessary

We use Principal Component Analysis (PCA) \parencite{hastie} on the SES distribution vectors and to identify the main axes of variation in the socioeconomic composition of individuals’ social networks. 
% PCA is a linear, unsupervised technique that transforms the original high-dimensional space (10 SES deciles in our case) into a smaller set of orthogonal components that capture the maximum variance in the data. 
We then assess the obtained principal components based on the share of explained variance. We select the smallest number of components that account for a substantial proportion of total variance while remaining interpretable in terms of meaningful patterns in the socioeconomic composition of individuals' networks. 
We interpret the components by examining their factor loadings across the SES percentiles. Finally, we compute obtained principal component scores for each individual in the sample and use them as data-driven indicators of social capital structure. 
%With these scores, we examine how these latent dimensions of social capital relate to each other and predict SES similarly to the previous individual-level approach in our statistical models using the latent dimensions instead of the pre-defined forms of EC and diversity. % this is not a method explanation if you ask me
%\end{itemize}
%\section*{Acknowledgements}

%[TO BE ADDED]

\section*{Data availability}

The socio-demographic and social network data that support the findings of this study are available from Statistics Netherlands (CBS). The population-scale network files, as well as individual-level socio-demographic information from 2009 to 2022 are included in the Statistics Netherlands Microdata Catalogue and are available for access through the Statistics Netherlands Microdata Services Remote Access Environment (RA). Under certain conditions, these microdata are accessible for scientific research. For further information, visit \cite{cbs_microdata}.

\printbibliography
%\vspace{180mm} 
%\break
%Place this at the end of your document to run TeXcount and filter by section names
%\immediate\write18{texcount -inc -incbib -sum -1 -sub=section main.tex > full_count.txt}

%\immediate\write18{texcount -inc -incbib -sub=section main.tex | grep "Introduction" | cut -d'+' -f1 > intro_count.txt}

%\immediate\write18{texcount -inc -incbib -sub=section main.tex | grep "Results" | cut -d'+' -f1 > results_count.txt}

%\immediate\write18{texcount -inc -incbib -sub=section main.tex | grep "Methods" | cut -d'+' -f1 > methods_count.txt}

%\immediate\write18{texcount -inc -incbib -sub=section main.tex | grep "Discussion" | cut -d'+' -f1 > discussion_count.txt}

%\section*{Word Count}
%\textbf{Introduction Word Count:}
%\verbatiminput{intro_count.txt}
%\textbf{Results Word Count:}
%\verbatiminput{results_count.txt}
%\textbf{Discussion Word Count:}
%\verbatiminput{discussion_count.txt}

%\textbf{Methods Word Count:}
%\verbatiminput{methods_count.txt}
%\textbf{All Text Word Count:}
%\verbatiminput{full_count.txt}

%\immediate\write18{texcount -inc -incbib -sub=section main.tex | grep "Abstract" | cut -d'+' -f1 > abstract_count.txt}

\clearpage
\section* {Supplementary Information}
\beginsupplement

\begin{table}[htbp]
\centering
\footnotesize
\resizebox{\textwidth}{!}{%
\begin{tabular}{lcccccc}
%\toprule
\hline 
%\hline \\[-1.8ex] 
& \multicolumn{6}{c}{\textit{Dependent variable:}} \\
%\cline{2-7} 
& \multicolumn{6}{c}{Future SES} \\
 & \multicolumn{2}{c}{$t_{1y}$} & \multicolumn{2}{c}{$t_{5y}$} & \multicolumn{2}{c}{$t_{10y}$} \\
\cmidrule(lr){2-3} \cmidrule(lr){4-5} \cmidrule(lr){6-7}
 & \shortstack{Socio-demographic \\characteristics (SD)} & \shortstack{SD + \\Parental SES} & SD &  \shortstack{SD + \\Parental SES} & SD &  \shortstack{SD + \\Parental SES}\\
\midrule
Parental SES & & $0.651^{***}$ & & $0.536^{***}$ & & $0.459^{***}$ \\
 & & (0.002) & & (0.003) & & (0.009) \\
Age & $-0.481^{***}$ & 0.020 & $-1.149^{***}$ & $-0.613^{***}$ & $-0.988^{***}$ & $-0.580^{***}$ \\
 & (0.026) & (0.017) & (0.037) & (0.030) & (0.111) & (0.096) \\
Sex & $-2.154^{***}$ & $-1.732^{***}$ & $-2.270^{***}$ & $-1.916^{***}$ & $-2.102^{***}$ & $-0.993^{*}$ \\
 & (0.156) & (0.104) & (0.194) & (0.155) & (0.658) & (0.568) \\
1\textsuperscript{st}-gen. migrant & $-22.779^{***}$ & $-5.625^{***}$ & $-20.996^{***}$ & $-6.723^{***}$ & $-20.153^{***}$ & $-7.178^{***}$ \\
 & (0.321) & (0.220) & (0.402) & (0.332) & (1.249) & (1.114) \\
2\textsuperscript{nd}-gen. migrant & $-16.507^{***}$ & $-3.711^{***}$ & $-15.430^{***}$ & $-4.650^{***}$ & $-13.413^{***}$ & $-3.807^{***}$ \\
 & (0.170) & (0.119) & (0.211) & (0.178) & (0.680) & (0.618) \\
Higher education & $18.228^{***}$ & $7.772^{***}$ & $18.881^{***}$ & $9.822^{***}$ & $15.908^{***}$ & $8.677^{***}$ \\
 & (0.745) & (0.497) & (0.932) & (0.749) & (3.162) & (2.720) \\
Secondary education & $9.637^{***}$ & $3.707^{***}$ & $8.378^{***}$ & $3.812^{***}$ & $5.217^{*}$ & 1.083 \\
 & (0.730) & (0.487) & (0.918) & (0.737) & (3.111) & (2.673) \\
Constant & $73.086^{***}$ & $25.517^{***}$ & $90.487^{***}$ & $48.956^{***}$ & $89.157^{***}$ & $54.993^{***}$ \\
 & (0.938) & (0.640) & (1.216) & (1.001) & (3.803) & (3.344) \\
\midrule
Observations & 91,010 & 90,346 & 60,434 & 59,888 & 6,899 & 6,851 \\
R$^2$ & 0.156 &\textbf {0.626} & 0.163 & \textbf{0.467} & 0.123 & \textbf{0.351} \\
Adj. R$^2$ & 0.156 & 0.626 & 0.163 & 0.467 & 0.123 & 0.351 \\
Residual Std. Error & \shortstack{21.447 \\(df = 91003) }& \shortstack{14.199 \\(df = 90338) }& \shortstack{21.142 \\(df = 60427) }& \shortstack{16.826\\ (df = 59880)} &\shortstack{ 21.661\\ (df = 6892) }&\shortstack{ 18.601\\ (df = 6843)} \\
F-Statistic & \shortstack{2807.063$^{***}$ \\(df = 6; 91003)} & \shortstack{21588.690$^{***}$ \\(df = 7; 90338) }& \shortstack{1958.765$^{***}$ \\(df = 6; 60427)} & \shortstack{7498.408$^{***}$ \\(df = 7; 59880)} &\shortstack{ 161.760$^{***}$ \\(df = 6; 6892)} & \shortstack{529.784$^{***}$ \\(df = 7; 6843)} \\
\hline 
%\hline \\[-1.8ex] 
%\bottomrule
\multicolumn{7}{r}{\footnotesize Note: $^{*}p<0.1$; $^{**}p<0.05$; $^{***}p<0.01$} \\
\hline
\end{tabular}
}
\caption{Coefficients of multivariate regression predicting individual SES in years 1, 5, and 10 based on individual socio-demographic features and when controlling for parental SES before entering economic independence.}
\label{tab:table1}
\end{table}

\clearpage
\begin{footnotesize}
\begin{longtable}{clccccc}
\caption{Coefficients of multivariate regression predicting individual SES in years 1, 5, and 10 based on level of Economic Connectedness, individual socio-demographic features, and when controlling for parental SES before entering economic independence.}
\label{tab:table_sc1}\\
\toprule
Time  & Predictor & Lower & Lower-middle & Middle & Upper-middle & Upper \\
\midrule
\endfirsthead

\multicolumn{7}{c}{{\bfseries \tablename\ \thetable{} -- continued from previous page}} \\
\toprule
Time  & Predictor & Lower & Lower-middle & Middle & Upper-middle & Upper \\
\midrule
\endhead

\midrule \multicolumn{7}{r}{{Continued on next page}} \\
\endfoot

\bottomrule
\endlastfoot

% ------------------ Year t_1 ------------------
\multirow{22}{*}{$t_{1y}$} & EC & 13.140*** & 12.505*** & 13.747*** & 16.110*** & 28.107*** \\
                        &    & (1.594)   & (0.774)   & (0.620)   & (0.613)   & (0.786)   \\
                        & Degree & 0.018*** & -0.010*** & -0.005*** & -0.003*** & 0.0004 \\
                        &        & (0.002)  & (0.001)   & (0.001)   & (0.001)   & (0.001) \\
                        & Parental SES & -0.620*** & 0.833*** & 0.741*** & 0.614*** & 0.572*** \\
                        &        & (0.036)  & (0.018)   & (0.014)   & (0.013)   & (0.016) \\
                        & Age & -0.579*** & 0.172*** & 0.223*** & 0.160*** & 0.019 \\
                        &      & (0.073)   & (0.035)   & (0.029)   & (0.028)   & (0.035) \\
                        & Being a woman & -2.528*** & -2.416*** & -1.234*** & -1.173*** & -0.688*** \\
                        &                & (0.458)   & (0.217)   & (0.178)   & (0.166)   & (0.203) \\
                        & 1\textsuperscript{st}-gen. migrant & -11.833*** & -1.049*** & -1.605*** & -1.049** & -1.460* \\
                        &                & (0.672)   & (0.362)   & (0.376)   & (0.482)   & (0.781) \\
                        & 2\textsuperscript{nd}-gen. migrant & -9.421*** & -0.293 & -0.306 & -0.238 & 0.297 \\
                        &                & (0.493)   & (0.230)   & (0.195)   & (0.210)   & (0.318) \\
                        & Higher education & 16.105*** & 7.285*** & 5.281*** & 5.091*** & -1.224 \\
                        &                & (1.917)   & (0.896)   & (0.809)   & (0.856)   & (1.234) \\
                        & Secondary education & 4.488** & 2.966*** & 1.814** & 3.162*** & -1.572 \\
                        &                & (1.861)   & (0.856)   & (0.783)   & (0.837)   & (1.222) \\
                        & Constant & 51.480*** & 7.771*** & 10.135*** & 16.470*** & 20.797*** \\
                        &          & (2.704)   & (1.353)   & (1.277)   & (1.413)   & (2.103) \\
\hline
&Observations & 9,726 & 14,244 & 20,015 & 22,989 & 23,226 \\
&R\textsuperscript{2} & 0.184 & 0.191 & 0.187 & 0.143 & 0.122 \\
&Adjusted R\textsuperscript{2} & 0.183 & 0.190 & 0.186 & 0.143 & 0.121 \\
&Residual Std. Error & \shortstack{20.912\\ (df = 9715)}  & \shortstack{11.964\\ (df = 14233)}  & \shortstack{11.403\\ (df = 20004)}  & \shortstack{11.254\\ (df = 22978)}  & \shortstack{13.835\\ (df = 23215)}  \\
&F-Statistic & \shortstack{218.786***\\ (df = 10; 9715)}  & \shortstack{336.087***\\ (df = 10; 14233)}  & \shortstack{458.684***\\ (df = 10; 20004)}  & \shortstack{384.699***\\ (df = 10; 22978)}  & \shortstack{321.449***\\ (df = 10; 23215)}  \\

% ------------------ Year t_5 ------------------
\hline
\multirow{22}{*}{$t_{5y}$} & EC & 27.711*** & 22.940*** & 25.223*** & 24.983*** & 15.451*** \\
                        &    & (2.178)   & (1.396)   & (1.094)   & (0.980)   & (0.958)   \\
                        & Degree & 0.016*** & -0.003 & -0.002 & 0.003*** & 0.001 \\
                        &        & (0.003)  & (0.002) & (0.002) & (0.001)  & (0.001) \\
                        & Parental SES & -0.662*** & 0.750*** & 0.502*** & 0.482*** & 0.677*** \\
                        &        & (0.050)  & (0.032) & (0.025) & (0.021)  & (0.020) \\
                        & Age & -0.621*** & -0.551*** & -0.622*** & -0.442*** & -0.373*** \\
                        &     & (0.114)   & (0.075)   & (0.061)   & (0.054)   & (0.051) \\
                        & Being a woman & -1.232* & -1.664*** & -2.141*** & -1.686*** & -1.665*** \\
                        &               & (0.648) & (0.402)   & (0.318)   & (0.268)   & (0.253) \\
                        & 1\textsuperscript{st}-gen. migrant & -10.022*** & -2.114*** & -2.835*** & -1.534** & -0.793 \\
                        &               & (0.940) & (0.663)   & (0.681)   & (0.780)   & (1.085) \\
                        & 2\textsuperscript{nd}-gen. migrant & -7.337*** & -0.563 & -0.715** & -0.750** & -1.450*** \\
                        &               & (0.691) & (0.422)   & (0.350)   & (0.346)   & (0.412) \\
                        & Higher education & 13.910*** & 9.857*** & 8.476*** & 3.781*** & 1.041 \\
                        &               & (2.820) & (1.605)   & (1.468)   & (1.349)   & (1.664) \\
                        & Secondary education & 1.816 & 2.964* & 2.735* & 0.153 & -0.457 \\
                        &               & (2.742) & (1.540)   & (1.429)   & (1.321)   & (1.650) \\
                        & Constant & 54.003*** & 27.351*** & 38.180*** & 37.276*** & 26.381*** \\
                        &          & (3.964)   & (2.595)   & (2.392)   & (2.328)   & (2.739) \\
\hline
&Observations & 5,857 & 9,390 & 13,619 & 15,784 & 15,143 \\
&R\textsuperscript{2} & 0.188 & 0.129 & 0.126 & 0.112 & 0.118 \\
&Adjusted R\textsuperscript{2} & 0.186 & 0.128 & 0.125 & 0.112 & 0.118 \\
&Residual Std. Error & \shortstack{22.471\\ (df = 5846)}  & \shortstack{17.465\\ (df = 9379)}  & \shortstack{16.304\\ (df = 13608)}  & \shortstack{14.690\\ (df = 15773)}  & \shortstack{13.533\\ (df = 15132)}  \\
&F-Statistic & \shortstack{135.094***\\ (df = 10; 5846)}  & \shortstack{138.428***\\ (df = 10; 9379)}  & \shortstack{196.269***\\ (df = 10; 13608)}  & \shortstack{199.134***\\ (df = 10; 15773)}  & \shortstack{203.413***\\ (df = 10; 15132)}  \\

\hline
% Add Year t_{10} here similarly...
\multirow{20}{*}{$t_{10y}$}         & EC                                                    & 42.788***  & 35.147*** & 37.961*** & 27.483*** & 13.804*** \\
                                   &                                                       & (6.366)    & (4.766)   & (3.596)   & (3.468)   & (3.014)   \\
                                   & Degree                                                & 0.016*     & 0.006     & 0.0004    & 0.005     & -0.004    \\
                                   &                                                       & (0.009)    & (0.007)   & (0.005)   & (0.004)   & (0.003)   \\
                                   %& Excess closure                                        & -11.365    & -17.939** & 7.369     & 6.072     & 5.415     \\
                                   %&                                                       & (11.047)   & (8.368)   & (6.307)   & (5.324)   & (3.498)   \\
                                   & Parental SES                                          & -0.600***  & 0.477***  & 0.427***  & 0.337***  & 0.700***  \\
                                   &                                                       & (0.154)    & (0.111)   & (0.082)   & (0.072)   & (0.061)   \\
%\multirow{16}{*}{$t_{10y}$}        
& Age                                                   & 0.094      & -0.736*** & -0.770*** & -0.382**  & -0.142    \\
                                   &                                                       & (0.343)    & (0.264)   & (0.200)   & (0.188)   & (0.149)   \\
                                   & Being a woman                                         & 1.349      & -2.161    & -0.232    & -0.373    & -1.055    \\
                                   &                                                       & (2.014)    & (1.531)   & (1.155)   & (1.082)   & (0.901)   \\
                                   & 1\textsuperscript{st}-gen. migrant & -8.229***  & -0.612    & -4.109*   & -1.298    & 1.981     \\
                                   &                                                       & (2.887)    & (2.378)   & (2.263)   & (3.350)   & (3.024)   \\
                                   & 2\textsuperscript{nd}-gen. migrant & -5.691***  & 1.605     & 3.898***  & -2.162*   & 1.172     \\
                                   &                                                       & (2.147)    & (1.593)   & (1.211)   & (1.289)   & (1.336)   \\
                                   & Higher education                                      & 13.342     & 0.990     & 8.923*    & 8.532     & 1.525     \\
                                   &                                                       & (8.937)    & (6.067)   & (4.894)   & (7.120)   & (4.836)   \\
                                   & Secondary education                                   & 0.008      & -7.590    & 2.956     & 3.436     & -2.022    \\
                                   &                                                       & (8.784)    & (5.874)   & (4.763)   & (7.047)   & (4.780)   \\
                                   & Constant                                              & 37.424***  & 49.111*** & 35.555*** & 38.316*** & 19.860**  \\
                                   &                                                       & (12.351)   & (9.283)   & (7.638)   & (9.751)   & (8.385)  \\
\hline
&Observations & 656 & 1,001 & 1,568 & 1,789 & 1,831 \\
&R\textsuperscript{2} & 0.219 & 0.118 & 0.145 & 0.082 & 0.110 \\
&Adjusted R\textsuperscript{2} & 0.206 & 0.109 & 0.140 & 0.077 & 0.105 \\
&Residual Std. Error & \shortstack{22.910\\ (df = 645)}  & \shortstack{20.054\\ (df = 990)}  & \shortstack{18.104\\ (df = 1557)}  & \shortstack{17.183\\ (df = 1778)}  & \shortstack{14.227\\ (df = 1820)}  \\
&F-Statistic & \shortstack{18.045***\\ (df = 10; 645)}  & \shortstack{13.206***\\ (df = 10; 990)}  & \shortstack{26.457***\\ (df = 10; 1557)}  & \shortstack{15.863***\\ (df = 10; 1778)}  & \shortstack{22.559***\\ (df = 10; 1820)}  \\
\hline 
%\hline \\[-1.8ex] 
\multicolumn{7}{r}{\footnotesize Note: $^{*}p<0.1$; $^{**}p<0.05$; $^{***}p<0.01$} \\

\end{longtable}
\end{footnotesize}
%DIVERSITY

\clearpage
\begin{footnotesize}
\begin{longtable}{clccccc}
\caption{Coefficients of multivariate regression predicting individual SES in years 1, 5, and 10 based on level of Socioeconomic Diversity, individual socio-demographic features, and when controlling for parental SES before entering economic independence.}
\label{tab:table_sc2} \\
\toprule
Time  & Variable & Lower & Lower-middle & Middle & Upper-middle & Upper \\
\midrule
\endfirsthead

\multicolumn{7}{c}{{\bfseries \tablename\ \thetable{} -- continued from previous page}} \\
\toprule
Time  & Variable & Lower & Lower-middle & Middle & Upper-middle & Upper \\
\midrule
\endhead

\midrule \multicolumn{7}{r}{{Continued on next page}} \\
\endfoot

\bottomrule
\endlastfoot

% Please add the following required packages to your document preamble:
% \usepackage{multirow}
\multirow{20}{*}{$t_{1y}$}  &  \shortstack[l]{Socioeconomic\\diversity}                               & 1.040***   & 0.138***  & -0.011    & -0.152*** & -0.592*** \\
                          &                                                       & (0.064)    & (0.035)   & (0.030)   & (0.027)   & (0.028)   \\
                          & Degree                                                & 0.008***   & -0.010*** & -0.006*** & -0.004*** & -0.002*   \\
                          &                                                       & (0.002)    & (0.001)   & (0.001)   & (0.001)   & (0.001)   \\
                          %& Excess closure                                        & -8.038***  & 5.893***  & 6.232***  & 6.809***  & 4.924***  \\
                          %&                                                       & (2.590)    & (1.373)   & (1.090)   & (0.958)   & (0.980)   \\
                          & Parental SES                                          & -0.509***  & 0.859***  & 0.795***  & 0.658***  & 0.703***  \\
                          &                                                       & (0.037)    & (0.018)   & (0.014)   & (0.013)   & (0.017)   \\
                          & Age                                                   & -0.599***  & 0.152***  & 0.194***  & 0.116***  & -0.037    \\
                          &                                                       & (0.072)    & (0.035)   & (0.029)   & (0.028)   & (0.036)   \\
                          & Being a woman                                         & -2.493***  & -2.498*** & -1.379*** & -1.367*** & -1.024*** \\
                          &                                                       & (0.454)    & (0.219)   & (0.180)   & (0.168)   & (0.206)   \\
                          & 1\textsuperscript{st}-gen. migrant & -11.908*** & -2.255*** & -3.163*** & -2.670*** & -2.901*** \\
                          &                                                       & (0.655)    & (0.358)   & (0.374)   & (0.485)   & (0.793)   \\
                          & 2\textsuperscript{nd}-gen. migrant & -9.473***  & -1.535*** & -1.674*** & -1.669*** & -0.949*** \\
                          &                                                       & (0.471)    & (0.219)   & (0.188)   & (0.205)   & (0.321)   \\
                          & Higher education                                      & 13.346***  & 8.634***  & 6.830***  & 6.921***  & 0.986     \\
                          &                                                       & (1.908)    & (0.904)   & (0.818)   & (0.867)   & (1.255)   \\
                          & Secondary education                                   & 3.806**    & 3.975***  & 2.934***  & 4.457***  & -0.046    \\
                          &                                                       & (1.842)    & (0.863)   & (0.791)   & (0.848)   & (1.242)   \\
                          & Constant                                              & 31.039***  & 8.439***  & 13.901*** & 25.871*** & 41.781*** \\
                          &                                                       & (3.044)    & (1.530)   & (1.420)   & (1.497)   & (2.134)   \\
\hline

&Observations & 9,726 & 14,244 & 20,015 & 22,989 & 23,226 \\
&R\textsuperscript{2} & 0.200 & 0.177 & 0.167 & 0.119 & 0.091 \\
&Adjusted R\textsuperscript{2} & 0.199 & 0.176 & 0.166 & 0.119 & 0.091 \\
&Residual Std. Error & \shortstack{20.707\\ (df = 9715)}  & \shortstack{12.067\\ (df = 14233)}  & \shortstack{11.542\\ (df = 20004)}  & \shortstack{11.413\\ (df = 22978)}  & \shortstack{14.072\\ (df = 23215)}  \\
&F-Statistic & \shortstack{242.472***\\ (df = 10; 9715)}  & \shortstack{306.262***\\ (df = 10; 14233)}  & \shortstack{399.784***\\ (df = 10; 20004)}  & \shortstack{310.149***\\ (df = 10; 22978)}  & \shortstack{233.430***\\ (df = 10; 23215)}  \\

\hline

\multirow{20}{*}{$t_{5y}$} & \shortstack[l]{Socioeconomic\\diversity}                                       & 1.210***   & 0.273***  & -0.069    & -0.065    & -0.150*** \\
                          &                                                       & (0.093)    & (0.066)   & (0.055)   & (0.044)   & (0.035)   \\
                          & Degree                                                & 0.007*     & -0.006**  & -0.004**  & -0.002    & -0.002**  \\
                          &                                                       & (0.003)    & (0.002)   & (0.002)   & (0.001)   & (0.001)   \\
                          %& Excess closure                                        & -1.167     & 5.591**   & 9.110***  & 7.709***  & 4.757***  \\
                          %&                                                       & (3.675)    & (2.461)   & (1.921)   & (1.516)   & (1.169)   \\
                          & Parental SES                                          & -0.569***  & 0.794***  & 0.601***  & 0.547***  & 0.729***  \\
                          &                                                       & (0.051)    & (0.032)   & (0.025)   & (0.021)   & (0.020)   \\
                          & Age                                                   & -0.702***  & -0.647*** & -0.774*** & -0.624*** & -0.473*** \\
                          &                                                       & (0.114)    & (0.076)   & (0.062)   & (0.054)   & (0.051)   \\
                          & Being a woman                                         & -1.193*    & -1.788*** & -2.448*** & -1.980*** & -1.883*** \\
                          &                                                       & (0.647)    & (0.407)   & (0.323)   & (0.273)   & (0.255)   \\
                          & 1\textsuperscript{st}-gen.   migrant & -11.430*** & -4.383*** & -5.761*** & -4.269*** & -2.029*   \\
                          &                                                       & (0.919)    & (0.657)   & (0.682)   & (0.789)   & (1.090)   \\
                          & 2\textsuperscript{nd}-gen.   migrant & -8.898***  & -2.935*** & -3.397*** & -3.138*** & -2.354*** \\
                          &                                                       & (0.658)    & (0.401)   & (0.337)   & (0.341)   & (0.411)   \\
                          & Higher education                                      & 12.498***  & 12.222*** & 11.940*** & 6.152***  & 2.164     \\
                          &                                                       & (2.827)    & (1.630)   & (1.494)   & (1.376)   & (1.677)   \\
                          & Secondary education                                   & 2.377      & 4.867***  & 5.081***  & 1.763     & 0.423     \\
                          &                                                       & (2.739)    & (1.557)   & (1.454)   & (1.346)   & (1.662)   \\
                          & Constant                                              & 35.776***  & 29.689*** & 48.100*** & 51.044*** & 36.965*** \\
                          &                                                       & (4.481)    & (2.944)   & (2.657)   & (2.471)   & (2.737)   \\
\hline
&Observations & 5,857 & 9,390 & 13,619 & 15,784 & 15,143 \\
&R\textsuperscript{2} & 0.189 & 0.105 & 0.092 & 0.076 & 0.104 \\
&Adjusted R\textsuperscript{2} & 0.187 & 0.104 & 0.091 & 0.075 & 0.104 \\
&Residual Std. Error & \shortstack{22.457\\ (df = 5846)}  & \shortstack{17.699\\ (df = 9379)}  & \shortstack{16.619\\ (df = 13608)}  & \shortstack{14.989\\ (df = 15773)}  & \shortstack{13.640\\ (df = 15132)}  \\
&F-Statistic & \shortstack{136.007***\\ (df = 10; 5846)}  & \shortstack{110.207***\\ (df = 10; 9379)}  & \shortstack{137.950***\\ (df = 10; 13608)}  & \shortstack{129.099***\\ (df = 10; 15773)}  & \shortstack{176.477***\\ (df = 10; 15132)}  \\

\hline

\multirow{20}{*}{$t_{10y}$} &  \shortstack[l]{Socioeconomic\\diversity}                                               & 1.313***   & 0.279     & 0.027     & -0.090    & 0.047     \\
                          &                                                       & (0.288)    & (0.242)   & (0.185)   & (0.156)   & (0.111)   \\
                          & Degree                                                & 0.006      & -0.002    & -0.006    & -0.002    & -0.009*** \\
                          &                                                       & (0.010)    & (0.007)   & (0.005)   & (0.005)   & (0.003)   \\
                         % & Excess closure                                        & -6.526     & -8.185    & 16.401**  & 13.474**  & 7.227**   \\
                         % &                                                       & (11.205)   & (8.481)   & (6.472)   & (5.352)   & (3.497)   \\
                          & Parental SES                                          & -0.525***  & 0.508***  & 0.583***  & 0.421***  & 0.724***  \\
                          &                                                       & (0.160)    & (0.114)   & (0.084)   & (0.072)   & (0.062)   \\
%\multirow{15}{*}{$t_{10y}$}   
& Age                                                   & -0.337     & -0.882*** & -1.088*** & -0.621*** & -0.328**  \\
                          &                                                       & (0.354)    & (0.271)   & (0.207)   & (0.192)   & (0.150)   \\
                          & Being a woman                                         & 1.374      & -2.203    & -0.966    & -0.737    & -1.285    \\
                          &                                                       & (2.051)    & (1.571)   & (1.195)   & (1.100)   & (0.906)   \\
                          & 1\textsuperscript{st}-gen.   migrant & -11.696*** & -5.883**  & -8.605*** & -5.141    & -0.180    \\
                          &                                                       & (2.852)    & (2.325)   & (2.301)   & (3.378)   & (3.012)   \\
                          & 2\textsuperscript{nd}-gen.   migrant & -9.217***  & -3.056**  & -0.683    & -4.642*** & 0.049     \\
                          &                                                       & (2.068)    & (1.497)   & (1.170)   & (1.273)   & (1.334)   \\
                          & Higher education                                      & 10.267     & 3.858     & 16.590*** & 9.156     & 1.602     \\
                          &                                                       & (9.145)    & (6.239)   & (5.022)   & (7.246)   & (4.864)   \\
                          & Secondary education                                   & -1.873     & -5.348    & 9.588*    & 3.480     & -1.903    \\
                          &                                                       & (8.955)    & (6.022)   & (4.890)   & (7.169)   & (4.809)   \\
                          & Constant                                              & 33.490**   & 58.325*** & 46.601*** & 55.565*** & 29.811*** \\
                          &                                                       & (13.372)   & (10.646)  & (8.479)   & (10.014)  & (8.181)   \\
\hline
&Observations & 656 & 1,001 & 1,568 & 1,789 & 1,831 \\
&R\textsuperscript{2} & 0.190 & 0.070 & 0.084 & 0.050 & 0.100 \\
&Adjusted R\textsuperscript{2} & 0.177 & 0.061 & 0.078 & 0.044 & 0.095 \\
&Residual Std. Error & \shortstack{23.326\\ (df = 645)}  & \shortstack{20.584\\ (df = 990)}  & \shortstack{18.740\\ (df = 1557)}  & \shortstack{17.483\\ (df = 1778)}  & \shortstack{14.309\\ (df = 1820)}  \\
&F-Statistic & \shortstack{15.126***\\ (df = 10; 645)}  & \shortstack{7.505***\\ (df = 10; 990)}  & \shortstack{14.291***\\ (df = 10; 1557)}  & \shortstack{9.291***\\ (df = 10; 1778)}  & \shortstack{20.248***\\ (df = 10; 1820)}  \\
\hline 
%\hline \\[-1.8ex] 
\multicolumn{7}{r}{\footnotesize Note: $^{*}p<0.1$; $^{**}p<0.05$; $^{***}p<0.01$} \\

\end{longtable}
\end{footnotesize}

\clearpage
\begin{footnotesize}
\begin{longtable}{clccccc}
\caption{Coefficients of multivariate regression predicting individual SES in years 1, 5, and 10 based on level of Excess Closure, individual socio-demographic features, and when controlling for parental SES before entering economic independence.}
\label{tab:table_sc3} \\
\toprule
Time  & Variable & Lower & Lower-middle & Middle & Upper-middle & Upper \\
\midrule
\endfirsthead

\multicolumn{7}{c}{{\bfseries \tablename\ \thetable{} -- continued from previous page}} \\
\toprule
Time  & Variable & Lower & Lower-middle & Middle & Upper-middle & Upper \\
\midrule
\endhead

\midrule \multicolumn{7}{r}{{Continued on next page}} \\
\endfoot

\bottomrule
\endlastfoot

% Please add the following required packages to your document preamble:
% \usepackage{multirow}

\multirow{20}{*}{$t_{1y}$}   
                           & Excess closure                                      & -6.921***  & 6.055***                    & 6.222***                    & 6.727***  & 4.504***  \\
                           &                                                     & (2.623)    & (1.373)                     & (1.089)                     & (0.959)   & (0.990)   \\
                            & Degree                                              & 0.021***   & -0.008***                   & -0.006***                   & -0.006*** & -0.009*** \\
                           &                                                     & (0.002)    & (0.001)                     & (0.001)                     & (0.001)   & (0.001)   \\
                           & Parental SES                                        & -0.622***  & 0.858***                    & 0.795***                    & 0.654***  & 0.648***  \\
                           &                                                     & (0.036)    & (0.018)                     & (0.014)                     & (0.013)   & (0.017)   \\
                           & Age                                                 & -0.610***  & 0.155***                    & 0.193***                    & 0.107***  & -0.100*** \\
                           &                                                     & (0.073)    & (0.035)                     & (0.029)                     & (0.028)   & (0.036)   \\
                           & Being a woman                                       & -2.543***  & -2.510***                   & -1.378***                   & -1.363*** & -1.076*** \\
                           &                                                     & (0.460)    & (0.219)                     & (0.180)                     & (0.168)   & (0.208)   \\
                           & 1\textsuperscript{st}-gen.   migrant & -12.919*** & -2.232***                   & -3.167***                   & -2.755*** & -3.417*** \\
                           &                                                     & (0.661)    & (0.358)                     & (0.374)                     & (0.485)   & (0.800)   \\
                           & 2\textsuperscript{nd}-gen.   migrant & -10.661*** & -1.549***                   & -1.677***                   & -1.754*** & -1.483*** \\
                           &                                                     & (0.471)    & (0.219)                     & (0.187)                     & (0.205)   & (0.323)   \\
                           & Higher education                                    & 17.269***  & 9.097***                    & 6.807***                    & 6.622***  & 0.161     \\
                           &                                                     & (1.918)    & (0.897)                     & (0.816)                     & (0.867)   & (1.267)   \\
                           & Secondary education                                 & 5.333***   & 4.236***                    & 2.925***                    & 4.370***  & -0.088    \\
                           &                                                     & (1.864)    & (0.861)                     & (0.791)                     & (0.848)   & (1.254)   \\
                           & Constant                                            & 56.338***  & 11.291***                   & 13.668***                   & 23.006*** & 34.242*** \\
                           &                                                     & (2.648)    & (1.348)                     & (1.282)                     & (1.412)   & (2.125)   \\

\hline
                                   &Observations&9,726&14,244&20,015&22,989&23,226\\
&R\textsuperscript{2}&0.178&0.176&0.167&0.118&0.073\\
&Adjusted R\textsuperscript{2}&0.177&0.176&0.166&0.117&0.073\\
&Residual Std. Error&\shortstack{20.984\\ (df = 9716)}&\shortstack{12.073\\ (df = 14234)}&\shortstack{11.542\\ (df = 20005)}&\shortstack{11.421\\ (df = 22979)}&\shortstack{14.211\\ (df = 23216)}\\
&F-Statistic&\shortstack{233.937***\\ (df = 9; 9716) }&\shortstack{338.233***\\ (df = 9; 14234)}&\shortstack{ 444.207***\\ (df = 9; 20005) }&\shortstack{340.520***\\ (df = 9; 22979) }&\shortstack{203.868***\\ (df = 9; 23216)}\\

\hline
%\hline \\[-1.8ex] 

\multirow{20}{*}{$t_{5y}$}    
                           & Excess closure                                      & 1.757      & 5.993**                     & 9.044***                    & 7.660***  & 4.563***  \\
                           &                                                     & (3.721)    & (2.462)                     & (1.920)                     & (1.516)   & (1.169)   \\
                           & Degree                                              & 0.018***   & -0.003                      & -0.005***                   & -0.003**  & -0.004*** \\
                           &                                                     & (0.003)    & (0.002)                     & (0.002)                     & (0.001)   & (0.001)   \\
                           & Parental SES                                        & -0.684***  & 0.791***                    & 0.601***                    & 0.545***  & 0.716***  \\
                           &                                                     & (0.051)    & (0.032)                     & (0.025)                     & (0.021)   & (0.020)   \\
                           & Age                                                 & -0.708***  & -0.644***                   & -0.778***                   & -0.629*** & -0.498*** \\
                           &                                                     & (0.116)    & (0.076)                     & (0.062)                     & (0.054)   & (0.050)   \\
                           & Being a woman                                       & -1.162*    & -1.823***                   & -2.448***                   & -1.984*** & -1.903*** \\
                           &                                                     & (0.656)    & (0.408)                     & (0.323)                     & (0.273)   & (0.255)   \\
                           & 1\textsuperscript{st}-gen.   migrant & -12.872*** & -4.382***                   & -5.774***                   & -4.306*** & -2.161**  \\
                           &                                                     & (0.926)    & (0.657)                     & (0.682)                     & (0.789)   & (1.091)   \\
                           & 2\textsuperscript{nd}-gen.   migrant & -10.381*** & -2.993***                   & -3.405***                   & -3.172*** & -2.485*** \\
                           &                                                     & (0.657)    & (0.401)                     & (0.336)                     & (0.340)   & (0.410)   \\
                           & Higher education                                    & 16.425***  & 13.169***                   & 11.779***                   & 6.042***  & 1.925     \\
                           &                                                     & (2.852)    & (1.615)                     & (1.489)                     & (1.374)   & (1.677)   \\
                           & Secondary education                                 & 3.422      & 5.297***                    & 5.026***                    & 1.760     & 0.433     \\
                           &                                                     & (2.777)    & (1.555)                     & (1.453)                     & (1.346)   & (1.663)   \\
                           & Constant                                            & 65.442***  & 35.544***                   & 46.692***                   & 49.810*** & 35.187*** \\
                           &                                                     & (3.913)    & (2.583)                     & (2.409)                     & (2.322)   & (2.707)   \\
\hline
&Observations& 5,857& 9,390& 13,619& 15,784& 15,143 \\

&R\textsuperscript{2} &0.165& 0.104& 0.092& 0.076& 0.103\\
&Adjusted R\textsuperscript{2} &0.164& 0.103& 0.091& 0.075& 0.103\\

& Residual Std. Error & \shortstack{ 22.779\\ (df = 5847)} & \shortstack{17.714\\ (df = 9380)} & \shortstack{16.619\\ (df = 13609)} & \shortstack{14.989\\ (df = 15774)} & \shortstack{13.648\\ (df = 15133) }\\

&F-Statistic& \shortstack{ 128.573***\\ (df = 9; 5847)} & \shortstack{ 120.345***\\ (df = 9; 9380)} & \shortstack{ 153.096***\\ (df = 9; 13609)} & \shortstack{143.196***\\ (df = 9; 15774)} & \shortstack{193.816***\\ (df = 9; 15133)}\\

\hline
\multirow{20}{*}{$t_{10y}$}
                           & Excess closure                                      & -3.928     & -7.688                      & 16.437**                    & 13.247**  & 7.295**   \\
                           &                                                     & (11.361)   & (8.472)                     & (6.466)                     & (5.336)   & (3.493)   \\
                            & Degree                                              & 0.016*     & -0.0004                     & -0.005                      & -0.003    & -0.009*** \\
                           &                                                     & (0.010)    & (0.007)                     & (0.005)                     & (0.004)   & (0.003)   \\& Parental SES                                        & -0.674***  & 0.506***                    & 0.583***                    & 0.419***  & 0.728***  \\
                           &                                                     & (0.159)    & (0.114)                     & (0.084)                     & (0.072)   & (0.061)   \\
%\multirow{14}{*}{$t_{10y}$} 
& Age                                                 & -0.046     & -0.861***                  & -1.083***                   & -0.644***  & -0.312**  \\
                           &                                                     & (0.354)    & (0.270)                     & (0.204)                     & (0.188)   & (0.145)   \\
                           & Being a woman                                       & 1.232      & -2.232                      & -0.959                      & -0.752    & -1.273    \\
                           &                                                     & (2.081)    & (1.571)                     & (1.193)                     & (1.099)   & (0.905)   \\
                           & 1\textsuperscript{st}-gen.   migrant & -13.986*** & -6.060***                 - & 8.611***                    & -5.273    & -0.102    \\
                           &                                                     & (2.850)    & (2.320)                     & (2.300)                     & (3.369)   & (3.006)   \\
                           & 2\textsuperscript{nd}-gen.   migrant & -11.116*** & -3.158**                    & -0.680                 -    & 4.678***  & 0.119     \\
                           &                                                     & (2.057)    & (1.495)                     & (1.170)                     & (1.271)   & (1.324)   \\
                           & Higher education                                    & 14.511     & 4.589                      & 16.641***                    & 9.049     & 1.620     \\
                           &                                                     & (9.235)    & (6.208)                     & (5.008)                     & (7.242)   & (4.863)   \\
                           & Secondary education                                 & 0.110      & -5.153                      & 9.612**                     & 3.469     & -1.941    \\
                           &                                                     & (9.080)    & (6.021)                     & (4.885)                     & (7.168)   & (4.807)   \\
                           & Constant                                            & 59.022***  & 64.320***                  & 47.081***                   & 54.158***  & 30.235*** \\
                           &                                                     & (12.327)   & (9.291)                     & (7.823)                     & (9.708)   & (8.118)   \\
\hline
&Observations&   656&    1,001&    1,568&1,789&1,831\\

&R\textsuperscript{2}&  0.164&   0.069&    0.084&0.049&0.100\\

&Adjusted R\textsuperscript{2}&   0.152&   0.061&    0.079&0.045&0.096\\

&Residual Std. Error   & \shortstack{ 23.680 \\(df = 646)}& \shortstack{ 20.587 \\(df = 991)}  & \shortstack{18.734\\ (df = 1558)}  & \shortstack{17.479\\ (df = 1779)} & \shortstack{14.305\\ (df = 1821) } \\

&F-Statistic&   \shortstack{ 14.069*** \\(df = 9; 646)} &\shortstack{8.189***\\ (df = 9; 991)} &\shortstack{15.887*** \\(df = 9; 1558)} &\shortstack{10.291*** \\(df = 9; 1779)} &\shortstack{22.488*** \\(df = 9; 1821)} \\
\hline 
%\hline \\[-1.8ex] 
\multicolumn{7}{r}{\footnotesize Note: $^{*}p<0.1$; $^{**}p<0.05$; $^{***}p<0.01$} \\

\end{longtable}
\end{footnotesize}

\end{document}